\documentclass[twocolumn,showpacs,preprintnumbers,amsmath,amssymb]{revtex4}

\input{psfig.sty}

\def\apj{{ApJ}}

\newcommand{\PRD}{{Phys. Rev. D}}
\newcommand{\PRL}{{Phys. Rev. Lett}}

\newcommand{\mpl}{{M_{\rm pl}}}

\newcommand{\be}{\begin{equation}}
\newcommand{\ee}{\end{equation}}
\newcommand{\bea}{\begin{eqnarray}}
\newcommand{\eea}{\end{eqnarray}}

\begin{document}
%opening
\title{WMAP-normalized Inflationary Model Predictions and the Search for Primordial Gravitational Waves with Direct Detection Experiments}

\author{Brett C. Friedman, Asantha Cooray}
\address{Center for Cosmology, Dept. of Physics \& Astronomy,  University of California, Irvine, CA 92697-4575}

\author{Alessandro Melchiorri}
\address{Dipartimento di Fisica ``G. Marconi'' and INFN, sezione
  di Roma, Universita' di Roma ``La Sapienza'', Ple Aldo Moro 5,
  00185, Roma, Italy.}

\begin{abstract}
In addition to density perturbations, inflationary models of the early universe generally predict a stochastic background of
gravitational waves or tensor fluctuations. By making use of the inflationary flow approach for single field models and 
fitting the models with Monte-Carlo techniques to cosmic microwave background (CMB) data from the {\it Wilkinson Microwave Anisotropy Probe} (WMAP),
 we discuss the expected properties of the gravitational wave background from inflation
at scales corresponding to direct detection experiments with laser interferometers in space.
We complement the Monte-Carlo numerical calculations by including predictions expected under
 several classes of analytical inflationary models.
We find that an improved version of {\it Big Bang Observer} (BBO-grand)
can be used to detect a  gravitational wave background at 0.1 Hz with a corresponding CMB
tensor-to-scalar ratio above 10$^{-4}$.
Even if the CMB tensor-to-scalar ratio were to be above 10$^{-2}$, 
we suggest that BBO-grand will be useful to study inflationary models  as the standard version of BBO, with a sensitivity
to a stochastic gravitational wave background $\Omega_{\rm GW}h^2 > 10^{-17}$,
 will only allow a marginal detection of the amplitude while 
leaving the tensor spectral index at 0.1 Hz unconstrained. Also, inflationary models 
with a large tensor-to-scalar ratio predict a substantial negative tensor spectral index 
such that the gravitational wave amplitude is damped at direct detection frequencies. 
We also discuss the extent to which CMB measurements can be used to predict the gravitational wave background  amplitude 
in a direct detection experiment and how any measurement
of the amplitude and the spectral tilt of the gravitational wave background at direct detection frequencies 
together with the CMB tensor-to-scalar ratio can be used to establish slow-roll inflation.
\end{abstract}


\maketitle

\section{Introduction}
\label{sec:intro}

The data from high precision cosmological experiments relating cosmic microwave
background (CMB) radiation \cite{Spergel} and large-scale structure mass distribution from
galaxy redshift surveys have clearly demonstrated that the inflationary paradigm \cite{Guth81} is the most 
preferred description of the origin of density perturbations. While expectations
are generally consistent with existing observations, we still lack a complete understanding
of underlying physics related to inflation including the shape and the amplitude of the
potential responsible for superluminal expansion.

The cosmological observations, so far, have been restricted to studying the density, or scalar, fluctuations in the
universe both through CMB and large-scale structure. In addition to scalar perturbations,
tensor fluctuations in the form of a primordial gravitational wave background are expected to
be generated during inflation \cite{Abbott84}. If detected, the amplitude of the gravitational
wave background provides a direct estimate of the potential height
of the inflaton field when relevant modes exit the horizon.
This background is now the focus of a large number of ongoing and upcoming CMB polarization experiments
since gravitational waves leave a distinct signature in the polarization pattern of CMB photons
in the form of  B- or curl-modes \cite{Kamionkowski:96}. The {\it Wilkinson Microwave Anisotropy Probe} (WMAP; \cite{Hinshaw:2006ia}) data have now constrained
the the tensor-to-scalar ratio, $r_{\rm CMB}$, denoting the tensor power spectrum amplitude divided by the amplitude of the
scalar perturbation power spectrum, to be below 0.38 at CMB scales (at 95\% confidence level \cite{Kinney:06}). 

The planned next generation NASA mission for high precision polarization observations ({\it CMBpol} or {\it Inflation Probe}) 
has now set a detection goal for $r_{\rm CMB}$ above 10$^{-2}$ \cite{Weiss}.
In the future, with an ambitious experiment,
this limit could be further improved, though confusion from known polarized foregrounds involving dust and synchrotron
will limit CMB studies  to a level above a few times  10$^{-4}$ \cite{foregrounds}.
While all-sky maps of polarized synchrotron emission are now available with low-frequency channels of WMAP,
the aforementioned uncertainty is due to the lack of all-sky  maps at frequencies
dominated by polarized dust. With data from Planck, it is very likely that we will have a more exact estimate on the limit to
which a tensor background can be probed with  next generation CMB polarization experiments.
It is now clear that the limiting factor for CMB observations is not the ultimate lensing confusion
that was previously discussed in the literature with a minimum detectable tensor-to-scalar ratio around 10$^{-5}$ \cite{Kesden}.

Beyond CMB, there is now a growing interest to directly detect the  relic gravitational wave background  from inflation 
with laser interferometers in space with test masses separated by a few thousand to 50,000 kilometers.
Plans for such studies include NASA's  {\it Big Bang Observer} (BBO) \cite{BBO} and the {\it DECI-hertz Interferometer Gravitational Wave
Observatory} (DECIGO) in Japan \cite{Seto:01}. Other studies that explore possibilities
for future gravitational wave observations beyond first generation {\it Laser Interferometer Space Antenna} can
 be found in Ref.~\cite{Crowder}.
Based on the expected foreground confusion and technological improvements, current concept studies aim the frequency regime between 
0.1 Hz  to a few Hz \cite{Coo:05}. Here we will use 0.1 Hz with a corresponding scale $k \sim 6.47 \times 10^{13}$ Mpc$^{-1}$
 as the fiducial frequency for direct detection experiments.
 As CMB measurements are sensitive to density and tensor perturbations at horizon-size scales,
at $k \sim 0.002$ Mpc$^{-1}$, the two detection techniques are roughly sensitive to a gravitational wave background
at wavelengths separated  by 16 orders of magnitude. 

As discussed in a variety of recent papers
\cite{Turner:96,Ungarelli:05,Smith:06,Smith,Efstathiou},  
the large lever arm produced by combining information at CMB and direct detection scales
is expected to pin down inflationary models better than data from any one of the two scales alone. 
Recent studies have discussed the expected signal for direct detections 
either through exact calculations using specific analytic models \cite{Smith:06} or
through the use of numerically generated Monte-Carlo models \cite{Smith,Efstathiou} under the inflationary flow equations
\cite{hoffman/turner:2001,kinney:2002}. While general predictions exist and the expected level of  the background is understood
in terms of the CMB tensor-to-scalar ratio, past studies have paid little attention to issues of connecting CMB information with
the gravitational wave background data at direct detection frequencies to improve inflationary models.  While the long lever arm
connecting the two scales has been advocated to help discern inflationary model space, as we discuss here,
such a large lever arm also complicates a simple analysis based on a power-law extrapolation of tensor and scalar power spectra
from CMB scales at $k=0.002$ Mpc$^{-1}$ to direct detection at $k=6.47 \times 10^{13}$ Mpc$^{-1}$.

Furthermore, while past studies have discussed the possibility for a direct detection of the inflationary background with future
laser interferometers \cite{Seto:06}, the sensitivity requirements for a direct detection experiment have not been fully characterized. 
We will try to determine how sensitive direct detection experiments will need to be and put these sensitivities in context with CMB
polarization measurements.  We also extend previous analysis
that only concentrated on the gravitational wave background amplitude \cite{Efstathiou} to also discuss the detectability of the
tensor spectral index at these frequencies.
Our discussion is similar in spirit to the
recent study in Ref.~\cite{Smith}, but we improve the analysis by considering both numerical as well as analytical \cite{Smith:06}
models of inflation, combining separate analyses in the literature into a single discussion.

The combination of the tensor amplitude and the tilt can be used to
validate inflationary models as these observables are expected to satisfy certain consistency relations.
Similar tests are discussed at CMB scales in the literature, though these relations are poorly constrained
due to the fact that tensor tilt is generally hard to measure at CMB scales \cite{Song}. 
As an application of our models, we extend the discussion in Smith et al. \cite{Smith}  to consider
the possibility to test the inflationary consistency relation. This is done in Smith et al. under the assumption that one can
use the tensor tilt from direct detection experiments as a proxy for the tensor tilt at CMB scales.
Using numerical models, we show that for general inflaton potentials the tensor tilt at $k=6.47 \times 10^{13}$ Mpc$^{-1}$
 differs from the tensor tilt at $k=0.002$ Mpc$^{-1}$ by more than a factor of ten.
 Furthermore, beyond simply using a direct detection experiment to confirm a detection of tensor modes at CMB scales and thereby
improve inflationary model selection, we also explore the usefulness of a highly sensitive direct detection experiment
that is also capable of measuring the tensor tilt at direct detection frequencies. Unfortunately, while
experiments such as ultimate-DECIGO can reach gravitational wave background
amplitudes with the corresponding CMB tensor-to-scalar ratio of 10$^{-6}$ and above,
 tensor tilt will remain unconstrained for most of the  models with CMB tensor-to-scalar ratios below 10$^{-3}$ and tilts with  magnitude such that 
$|n_T|<0.01$.

This paper is organized as follows: In the next Section, we outline our calculations that we use to
predict properties of the gravitational wave background from inflationary models. We consider two separate
approaches, one involving Monte-Carlo models of the inflationary flow equations, and the second using analytical families of
inflationary models that are consistent with observations at CMB scales. In Section~III, we discuss our results and comment on
the sensitivity requirements for a direct detection experiment to reach similar sensitivities as those expected with CMB
polarization observations. We conclude with a summary of our main results in Section~IV.

\section{Calculational Method}

\subsection{Monte-Carlo Reconstruction of Inflationary Flows}

To model the evolution of single field inflationary models, we study dynamics through the hierarchy of flow
equations involving the generalized ``Hubble Slow Roll'' (HSR)
parameters \cite{hoffman/turner:2001, kinney:2002,
easther/kinney:2003}.  The advantage of the inflationary flow method is that
one can study the  generic behavior of the inflaton field without assuming a particular shape for the potential.
We outline the main equations of this approach here and refer the reader to Refs.~\cite{ kinney:2002,easther/kinney:2003}
for further details.

To begin, we note that the motion of the scalar field in a cosmological background  is given by
$\ddot{\phi}+3H\dot{\phi}+V'(\phi) = 0$ where an overdot corresponds to the time derivative, a prime denotes the
derivative with respect to $\phi$ and $H\equiv(\dot{a}/a)$ is the Hubble parameter.
Assuming the inflaton field dominates the energy density, the Einstein field equations are
\begin{eqnarray}
H^2 &=& \frac{8\pi}{3\mpl^2}\left[V(\phi)+\frac{1}{2}\dot{\phi}^2\right] \nonumber \\
\left(\frac{\ddot{a}}{a}\right) &=& \frac{8\pi}{3\mpl^2}\left[V(\phi)-\dot{\phi}^2\right] \, .
\label{eqn:potential}
\end{eqnarray}
As discussed in Refs.~\cite{Lidsey}, the equations of motion can be written as
\begin{eqnarray}
\dot{\phi} &=&- \frac{\mpl^2}{4\pi} H'(\phi) \nonumber \\
\left[H'(\phi)\right]^2 &=&\frac{12\pi}{\mpl^2}H^2(\phi)-\frac{32\pi^2}{\mpl^4}V(\phi) \, , \label{eq:hj}
\end{eqnarray}
where the  Hubble parameter is now a function of the field $\phi$ instead of time under the assumption that $\phi$ varies monotonically with time.

As usual, dynamics are described in terms of the HSR parameters that are defined in terms of the derivatives of $H$ with respect to $\phi$ 
\cite{Liddle}:
\begin{eqnarray}
\epsilon(\phi) &\equiv& \frac{M^2_{\rm pl}}{4\pi}\frac{[H'(\phi)]^2}{H^2(\phi)} \nonumber \\
^{\ell}\lambda_H &\equiv& \left(\frac{M^2_{\rm pl}}{4\pi}\right)^\ell
  \frac{(H')^{\ell-1}}{H^\ell} \frac{d^{(\ell+1)} H}{d\phi^{(\ell+1)}} \, ,  \quad \quad (\ell \geq 1)
 \label{eq:hier}
\end{eqnarray}
and  for comparison with similar calculations in the literature \cite{hoffman/turner:2001, kinney:2002,
easther/kinney:2003,Efstathiou} we identify the slow roll parameters $^1\lambda_H$ and $^2\lambda_H$ as $\eta$ and $\xi$, respectively.
We also define $\sigma \equiv 2\eta-4\epsilon$. 

The trajectories of these HSR parameters are governed by a set of
coupled first order differential equations \cite{hoffman/turner:2001, kinney:2002}
written in terms of the number of e-folds, $N$, before the end of inflation,
with the convention that $N$ increases as one goes further back in time, as
\begin{eqnarray}
\frac{d\epsilon}{dN}&=&\epsilon (\sigma+2 \epsilon) \\
\frac{d\sigma}{dN}&=&-5\epsilon \sigma-12 \epsilon^2+2 \xi  \nonumber \\
\frac{d\left(^\ell \lambda_H \right)}{dN}&=&\left[\frac{\ell-1}{2} \sigma+(\ell-2)\epsilon\right]\left(^\ell\lambda_H\right)+\left(^{\ell+1}\lambda_H\right) \,  (\ell > 1). \nonumber
\end{eqnarray}
As written, the evolution of a given slow roll parameter depends on a slow roll parameter that is one order higher.
While the coupled differential equations involve an infinite hierarchy, in practice, this hierarchy must be truncated at some order M such that
$^{M+1}\lambda_H = 0$. From Eq.~\ref{eq:hier}, with $d^{(M+2)} H/d\phi^{(M+2)} = 0$,
one can then write $H(\phi)$ as  a polynomial of order $M+1$ with \cite{liddle:2003}
\begin{equation}
H(\phi) = H_0\left[ 1+ A_1 \left(\frac{\phi}{\mpl}\right) + ... + A_{M+1}
\left(\frac{\phi}{\mpl}\right)^{M+1}\right]. \label{eq:h}
\end{equation}
Further, from the definition of $\epsilon(\phi)$, one can also write \cite{liddle:2003}
\begin{eqnarray}
&&\epsilon(\phi) = \frac{M^2_{\rm pl}}{4\pi} \\
&\times& \left[\frac{\left(A_1/\mpl\right) + ... + (M+1)
\left(A_{M+1}/\mpl\right)\left(\phi/\mpl\right)^M}{1+A_1\left(\phi/\mpl\right)
+ ... + A_{M+1}\left(\phi/\mpl\right)^{M+1}}\right]^2, \nonumber  \label{eq:epsanalytic}
\end{eqnarray}
when the coefficients $A_i$, with $i > 1$, are related to initial values of HSR parameters
\begin{eqnarray}
A_{\ell+1} &=& \frac{(4\pi)^\ell \ ^{\ell}\lambda_{H,0}}{(\ell+1)!
\ A_1^{\ell-1}} \, , \label{eq:coeffs}
\end{eqnarray}
with $ A_1 = \sqrt{4\pi\epsilon_0}$ and the sign of $A_1$ determining the direction the field is rolling.

Once $\epsilon(N)$ is determined as a function of the number of e-folds before the end of inflation $N$,
the relation between number of e-folds and the field $\phi$ is determined from
\begin{equation}
\frac{d\phi}{d N } = \frac{\mpl}{2\sqrt{\pi}}\sqrt{\epsilon} \, , \label{eq:Neq}
\end{equation}
while the Hubble parameter is established from
\begin{equation}
\frac{1}{H}\frac{dH}{dN}=\epsilon \, .
\end{equation}
With $\epsilon(\phi)$ and $H(\phi)$ reconstructed, one can also obtain the inflaton potential 
through the Hamilton-Jacobi equation
\begin{equation}
V(\phi) =
\left(\frac{3\mpl^2 H^2\left(\phi\right)}{8\pi}\right)
\left[1-\frac{1}{3}\epsilon\left(\phi\right)\right] \, . \label{eq:v}
\end{equation}
Using equation~(10) and the expressions in (\ref{eqn:potential}), one can write
$(\ddot{a}/a)=H^2(\phi)[1-\epsilon(\phi)]$. Inflation is defined as $(\ddot{a}/a)>0$, which happens as long as $\epsilon(\phi) < 1$.
We use this criteria in each of the numerically generated models, such that the end of inflation occurs when $\epsilon(\phi) \rightarrow 1$.

Note that there is an overall constant of $H_0$ here that is not set by the differential equations.
This constant can be established based on the observed amplitude of either tensor or scalar
fluctuations at a specific scale. Here, we will normalize our models based on the amplitude of
density fluctuations at scales corresponding to CMB observations as determined by WMAP data.
Since models are normalized to CMB, we pick the fiducial physical scale that corresponds to 
$\phi_{CMB}=0$  to match the usual pivot point of CMB observables at
$k_{\rm CMB}=0.002$ Mpc$^{-1}$. Beyond this pivot point,
the physical wavenumber is associated with a value of $\phi$ through
\begin{equation}
\frac{d\phi}{d\ln k} = -\frac{\mpl}{2\sqrt{\pi}}
\frac{\sqrt{\epsilon}}{1-\epsilon}\, . \label{eq:phieq}
\end{equation}
Our convention is such that $\phi>0$ corresponds to scales smaller than $k_{\rm CMB}$ ($k>0.002$ Mpc$^{-1}$) such that one is going ahead in time.

At a given wavenumber,  the observable power spectra of tensor and scalar perturbations are given as \cite{stewart/lyth:1993}
\begin{eqnarray}
P_T(k)&=& \frac{16}{\pi}\left[1-\left(\chi+1\right)\epsilon\right]^2 \left(\frac{H}{\mpl}\right)^2\Big|_{k=aH} \nonumber \\
P_S(k)&=& \frac{\left[1-\left(2\chi+1\right)\epsilon+\chi\eta\right]^2}{\pi\epsilon} \left(\frac{H}{\mpl}\right)^2\Big|_{k=aH} \, ,
\end{eqnarray}
when $\chi=-2+\ln 2+\gamma$ and $\gamma=0.5772156649$ is the Euler-Mascheroni Constant.
This allows us to set the overall normalization $H_0$ such that at $k_{\rm CMB}$ we take 
\begin{equation}
P_S(k=0.002)=(2.45 \pm 0.23)\times 10^{-9} \, ,
\end{equation}
consistent with WMAP \cite{Spergel}.
This normalization is equivalent to setting: 
\begin{equation}
\frac{H(\phi_{\rm CMB})}{\mpl}\frac{[1-(2\chi+1)\epsilon(\phi_{\rm CMB})+\chi\eta(\phi_{\rm CMB})]}{\sqrt{\pi \epsilon(\phi_{\rm CMB})}} \approx 4.9 \times 10^{-5} \, ,
\end{equation}
and for most practical purposes the term within the square bracket can be ignored 
\footnote{This WMAP-based normalization is higher than usual normalizations in the literature
with $H/\mpl\sqrt{\pi \epsilon} =10^{-5}$ \cite{Kinney:06} and generally results in a factor of a few to a ten difference between
various predictions in the literature related to the amplitude of the gravitational wave background at 0.1 Hz.}. With our choice that CMB scale
is at $\phi=0$, in equation~(5), $H_0=H_{\rm CMB}$.

Instead of amplitude at power-spectra at each $k$, the observables are generally described in terms of the power-law variables given as
\begin{eqnarray}
&&n_S=1+\sigma-(5-3C)\epsilon^2-\frac{1}{4} (3-5C)\sigma \epsilon + \frac{1}{2} (3-C)\xi \nonumber \\
&&n_T=-2\epsilon-(1-C)\epsilon^2+\frac{1}{2}(1+C)\epsilon\sigma \nonumber \\
&&\alpha_S \equiv \frac{dn_S}{dlnk} = -\frac{1}{4} \left(\frac{1}{1-\epsilon}\right) \nonumber \\
&&\times \Big[2(C-3) \left(^3\lambda_H\right) -2 \xi \left[4+(5C-3)\epsilon-(C-3)\sigma\right]\nonumber \\
&&+\epsilon[4(11+3C)\epsilon^2+\epsilon[48+(31-9C)\sigma] \nonumber \\
&&+\sigma(20+(3-5C)\sigma)]\Big] \nonumber \\
&&\alpha_T \equiv \frac{dn_T}{dlnk}=\frac{1}{1-\epsilon}\nonumber \\
&&\times \Big[2(\epsilon\sigma+2\epsilon^2)2\epsilon(1-C)(\epsilon\sigma+2\epsilon^2) \nonumber \\
&&-\frac{1}{2}\left(1+C\right)\left(-3\epsilon^2\sigma-\frac{1}{2}\epsilon^3+2\epsilon({}^2\lambda_H)+\epsilon\sigma^2\right) \Big] \, , 
\end{eqnarray}
where $C=4(\ln 2 + \gamma)-5=0.08145$. 
For simplicity, we also discuss the tensor-to-scalar ratio defined at a particular scale as
\begin{equation}
r \equiv\frac{P_T(k)}{P_S(k)} \,
\end{equation}
which in terms of slow-roll parameters is
\begin{equation}
r=16\epsilon\left[1-C(\sigma+2 \epsilon)\right] \, .
\end{equation}
To the first order in slow roll parameters, $r\approx-8n_t$, which is considered to be a consistency relation that can be used to validate
inflationary models. We will discuss the accuracy to which this relation is valid at both CMB and direct detection  scales.
Though we can only expect a measurement of the tensor amplitude at 0.1 Hz with direct detection experiments,
we will discuss the relation between tensor-to-scalar ratio at CMB scales $r_{\rm CMB}$ and the same ratio at direct detection frequencies, $r_{\rm 0.1 Hz}$.
Due to the scale dependence of the scalar and tensor tilts, these two ratios are not expected to be equal for a general inflationary potential.

To generate a family of models, we follow the usual recipes outlined in the literature \cite{kinney:2002,
easther/kinney:2003,Efstathiou}.  We take the initial conditions to be sampled with in the parameter range of
\begin{eqnarray}
\epsilon_0 &\in&[0,0.8] \nonumber \\
\sigma_0&\in&[-0.5,0.5] \nonumber \\
\xi_0&\in&[-0.05,0.05] \nonumber \\
^\ell\lambda_{H|0}&\in&[-0.025\times 5^{-\ell+3},0.025\times 5^{-\ell+3}], (3\leq \ell \leq 10) \nonumber \\
^{11}\lambda_{H|0}&=&0 \nonumber \, ,
\end{eqnarray}
such that the hierarchy is truncated at the tenth order in HSR parameters. 
We solve the flow equations
for each  set of numbers randomly drawn within the above ranges. Assuming each of the ranges is uniform, 
we establish $[\epsilon(N),^\ell \lambda_H(N)]$. We require that inflation last between $46$ and $67$ e-folds, roughly
consistent with known constraints \cite{Alabidi,Dodelson}.
With $\epsilon(N)$, we also solve for the relation between $\phi$ and $N$ and between $\phi$ and $k$. 

The latter establishes primordial power spectra as a function of $k$. The relations also establish
the power-law observables at a given wave number, though to compare with observations we restrict them
to the values at CMB scale, with $k_{\rm CMB}=0.002$ Mpc$^{-1}$, and $k_{\rm dir} = 6.47 \times 10^{13}$ Mpc$^{-1}$.
The large number of e-folds between these two scales, $\Delta N=\ln(k_{\rm dir}/k_{\rm CMB}) \sim 38$, is expected to improve
constraints on the inflationary model since the potential will be probed at two largely different field values \cite{Smith:06}.

Using the above initial conditions, we ran close to 20 million individual models \footnote{On a standard desktop machine, the 20 million Monte-Carlo models
took about 5 days to complete. The C code to generate model predictions related to tensor and scalar power spectra using the inflationary flow equations 
is freely available from the authors.}. Out of these 20 million models, roughly 20,000 models are
found to fall within WMAP 95\% confidence level constraints of $r_{\rm
  CMB} < 0.38$ and $0.92 <n_s<1.06$. 
The latter are suggested from recent cosmological data analyses (see
e.g. \cite{Spergel}, \cite{Kinney:06}) and broadly describe the
allowed parameter ranges within the 2$\sigma$ confidence level. 
In Fig.~1, we summarize the ($n_s,r$) plane and the distribution of
model points. We will discuss these and other results in Section~III.

\begin{figure*}[!t]
\centerline{\psfig{file=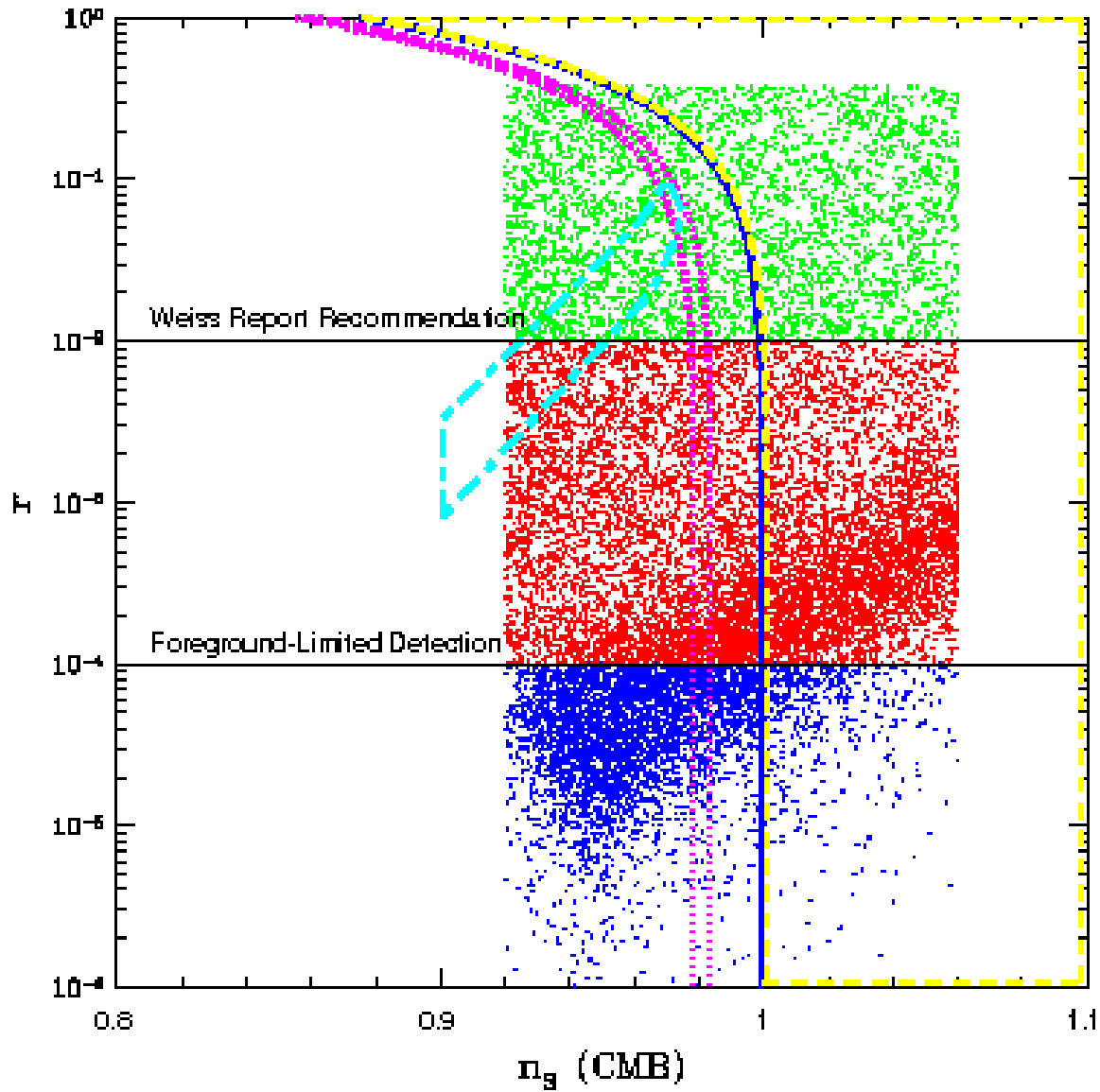,width=6.0in,angle=0}}
\caption{The region in ($n_s,r$) plane consistent with WMAP CMB data at the 2$\sigma$ confidence level with $r_{\rm CMB} < 0.38$ and $0.92 < n_s < 1.06$,
where $r_{\rm CMB}$ is the tensor-to-scalar ratio and $n_s$ is the scalar spectral index, or tilt. Both these parameters are determined
at CMB scales with $k=0.002$ Mpc$^{-1}$. Inflationary models corresponding to each  model point shown here also  give rise to
$N> 46$ e-folds in the scale factor  before the end of inflation.  For comparison, we also plot the parameter space occupied by
four models of inflation outlined in Section~IIB: power-law (solid blue line), chaotic (dotted magenta), symmetry breaking (dot-dashed cyan), and
hybrid (dashed yellow) models. The two horizontal lines outline the expectation on the tensor-to-scalar ratio from CMB studies. The two values
are chosen to the target goal of the next generation CMB polarization experiment ($r_{\rm CMB}>10^{-2}$) and an optimistic limit related
to polarized foreground confusion ($r_{\rm CMB} > 10^{-4}$). A more realistic estimate on the foreground confusion will suggest that CMB studies
will be limited anywhere between a tensor-to-scalar ratio of $10^{-3}$ to $10^{-4}$.
}
\end{figure*}

\begin{figure*}[!t]
\centerline{\psfig{file=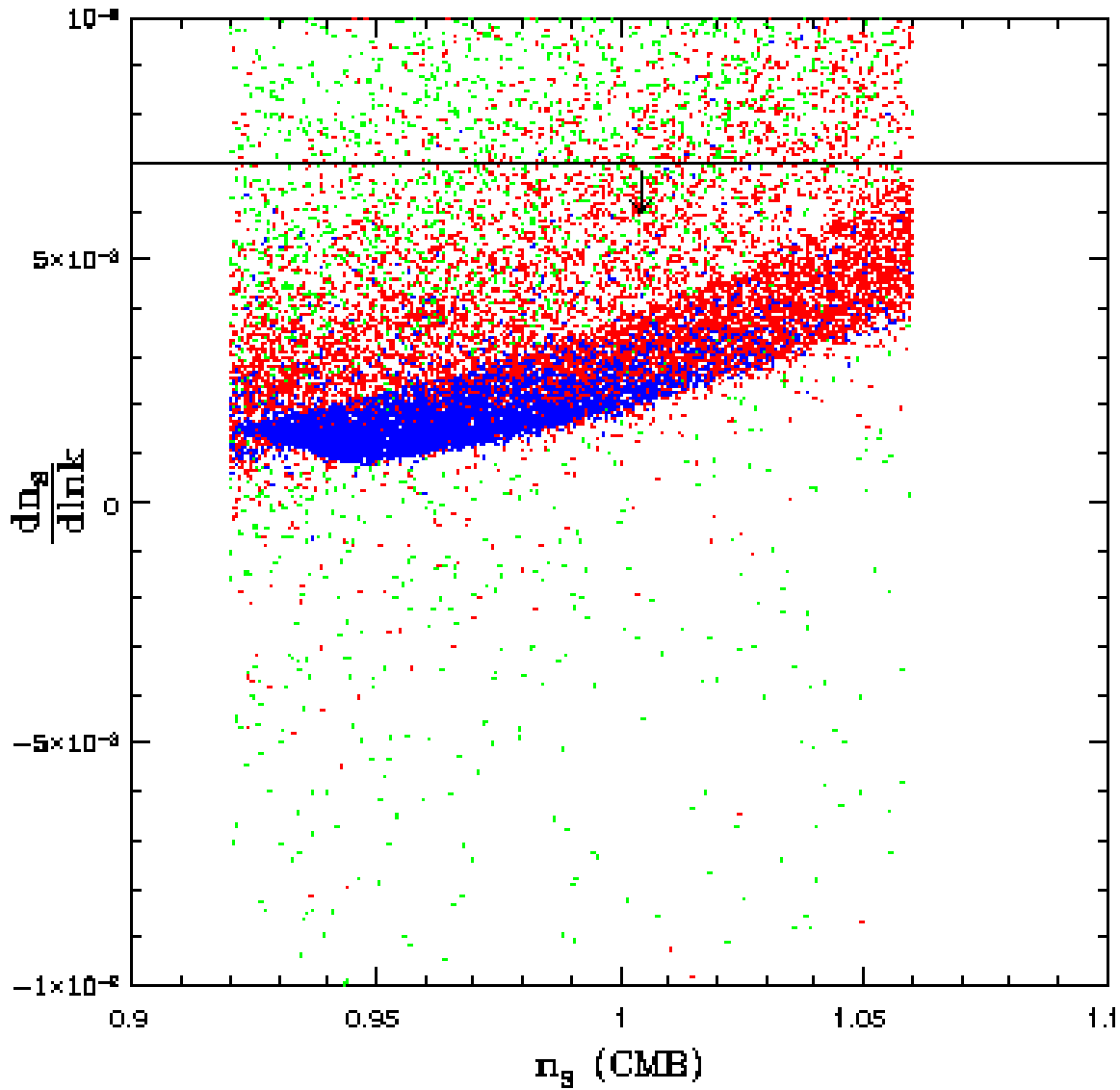,width=6.0in,angle=0}}
\caption{The running of the scalar spectral index against the spectral index at CMB scales. We did not constrain the running, although
a majority of the models fall below $10^{-2}$ and have positive running. The horizontal line at $\frac{dn_S}{dlnk}=0.007$ shows the WMAP 2$\sigma$ upper-limit
on the running of the scalar spectral index.
}
\end{figure*}

\begin{figure*}[!t]
\centerline{\psfig{file=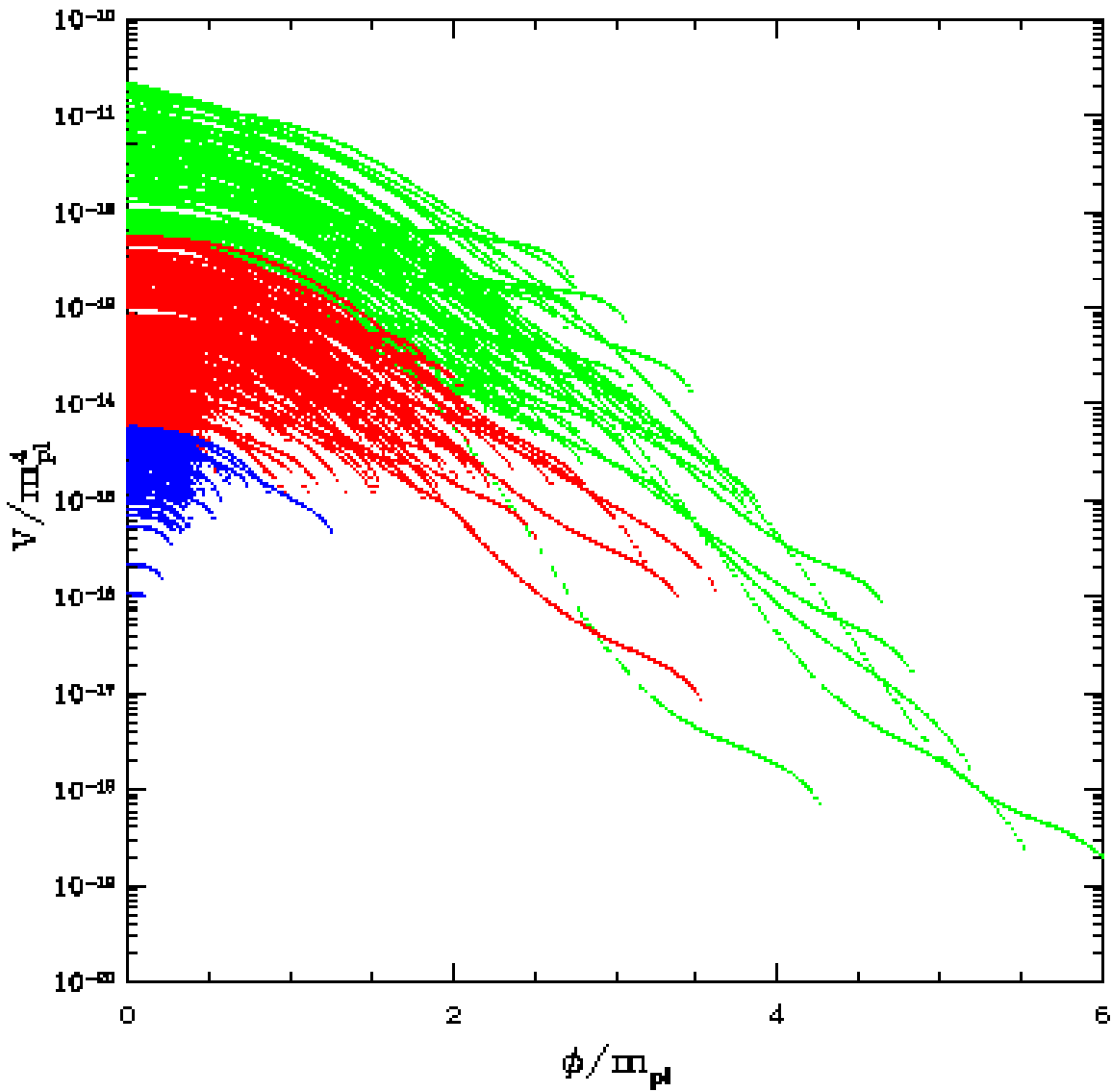,width=6.0in,angle=0}}
\caption{The set of potentials that give rise to points shown in Figure~1 from the Monte-Carlo inflationary flow analysis. 
Here $\phi=0$ corresponds to observations at CMB scales with $k=0.002$ Mpc$^{-1}$, while $\phi > 0$ corresponds to scales smaller
than CMB. 
The potentials are subdivided based tensor-to-scalar ratio at CMB. Potentials that give rise to large tensor-to-scalar ratios with $r_{\rm CMB} > 10^{-2}$
cannot simply be described by simple power-laws or any particular shape. 
}
\end{figure*}

\begin{figure*}[!t]
\centerline{\psfig{file=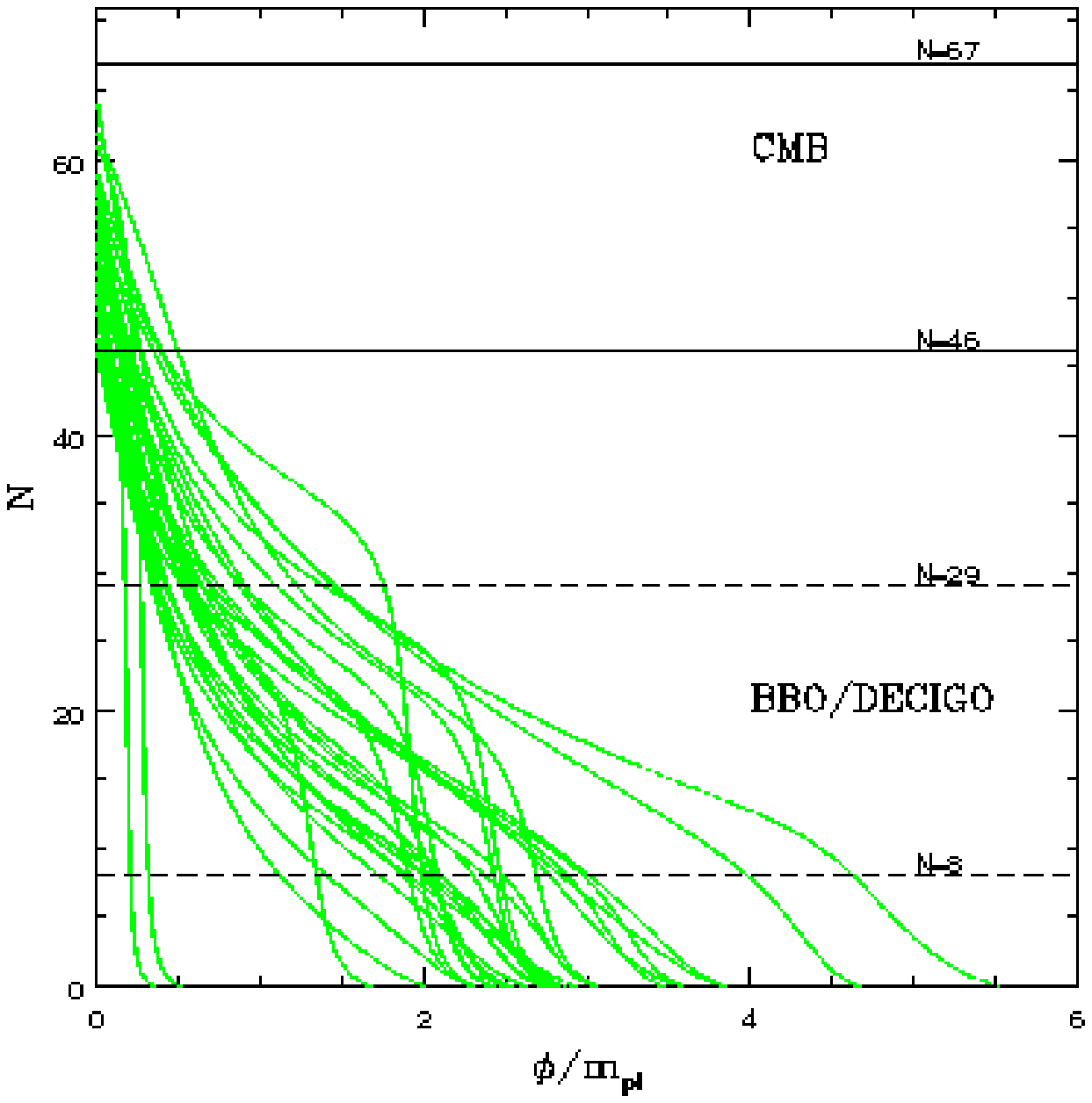,width=3.5in,angle=0}
\psfig{file=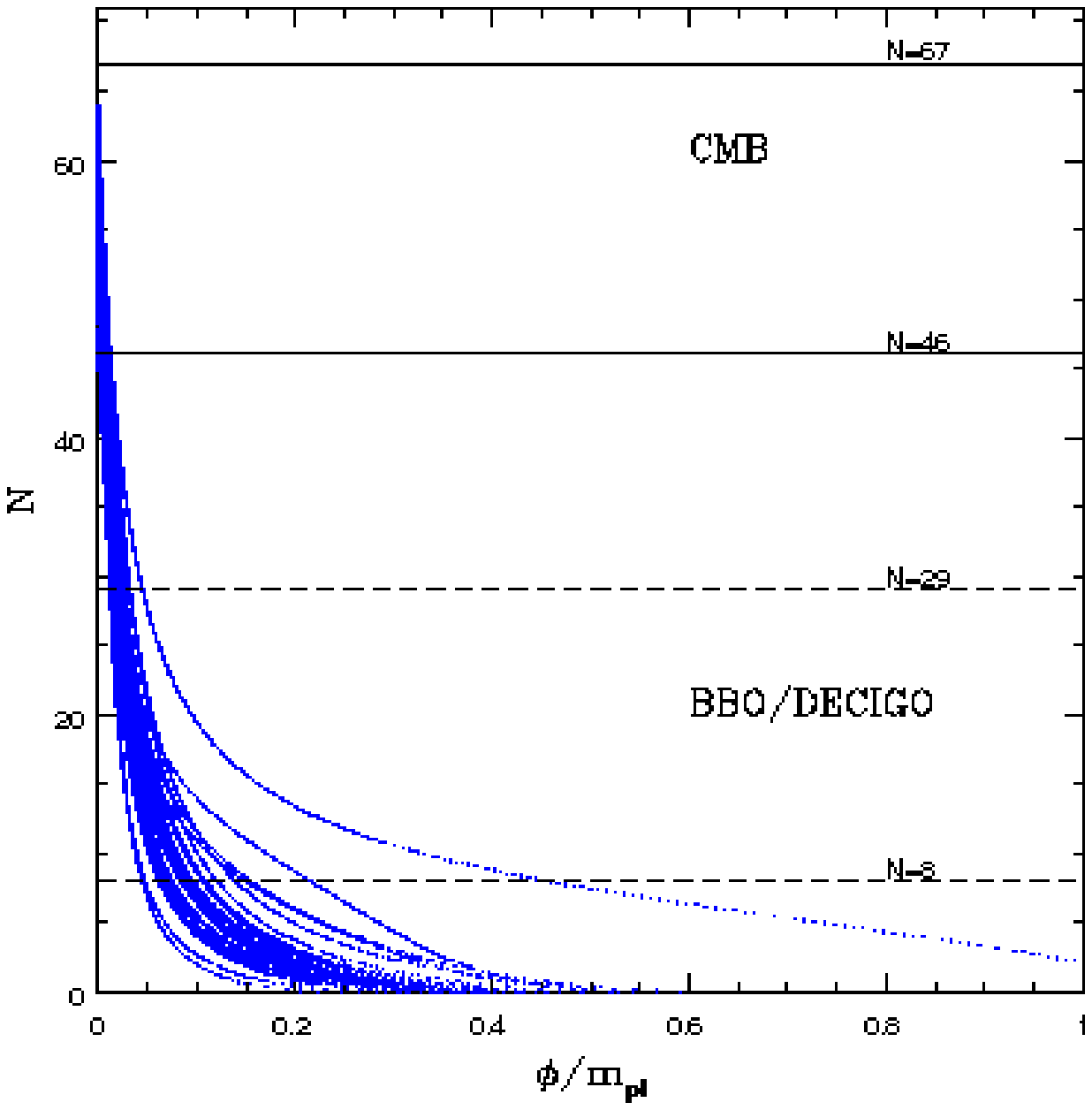,width=3.5in,angle=0}}
\caption{The relation between the number of e-folds $N$ and $\phi$, the value of the inflaton field, for representative sampling of potentials
shown in Figure~3.  In the left panel, we make use of potentials that give rise to $r_{\rm CMB} > 10^{-2}$, while in the right panel,
we concentrate on the relation between $N$ and $\phi$ for potentials that lead to $r_{\rm CMB} < 10^{-4}$. In the case of potentials that  
lead to large $r_{\rm CMB}$, $V(\phi)$'s are shaped such that there is a significant, and an abrupt, increase in the number of e-folds
over a small variation in $\phi$. The $\phi$ values related to this abrupt changes in $N$ correspond to the field values where the inflaton potentials in  
Figure~3 flatten. On the other hand, $N$ versus $\phi$ relations for potentials that results in  a small $r_{\rm CMB}$ are gradual. A behavior similar to this
was already described in Ref.~\cite{Efstathiou} based on the relation between $H(N)$ versus $N$
such that models that lead to large tensor amplitude have abrupt changes in $H(N)$. The horizontal lines indicate the
number of e-foldings related to CMB (top two lines) and direct detection (bottom two lines) observations.}
\end{figure*}

\subsection{Analytical Models of Slow-Roll Inflation}

In addition to the numerical method to describe inflationary predictions, we also make use of several analytical
models of slow-roll inflation to connect CMB observables with those at direct detection experiment levels. Unlike the previous
subsection where flow equations allow us to describe the dynamics without specifying a particular shape for $V(\phi)$,
with analytical models, we are required to specify the potential explicitly. For easy comparison,
we take the same four families of models as those considered in Ref.~\cite{Smith:06}. These models describe both single field models
and hybrid inflation models. The four potentials considered here are
\begin{eqnarray}
V(\phi)&=&V_0 e^{-p\phi/\mpl} \, , \quad \quad {\rm Power-law}\nonumber \\
V(\phi)&=&V_0 \left(\frac{\phi}{\mpl}\right)^\alpha \, \quad \quad {\rm Chaotic} \nonumber \\
V(\phi)&=&V_0 \left[1-\left(\frac{\phi}{\nu}\right)^2\right]^2 \,
\quad {\rm Symmetry-breaking}\nonumber \\
V(\phi)&=&V_0 \left[1+\left(\frac{\phi}{\mu}\right)^2\right]  \, , \quad {\rm Hybrid}
\end{eqnarray}
and describe power-law, chaotic, symmetric breaking, and hybrid inflation, respectively from top to bottom. 
The calculation involves simply matching the observables at CMB scales $(n_s,r)$
to free parameters in each of the potentials, such as $\alpha$ and $\nu$, and also uses the number of e-folds between CMB horizon exit and the end of
inflation. The amplitude of density perturbations $P_S(k_{\rm CMB})$ fixes the overall potential normalization given
by $V_0$. With $V_0$ and free parameters fixed, and taking $\Delta N=38$,  we calculate the observables in terms of $\Omega_{\rm GW}h^2$ (see below) 
and $n_T$ and, in cases where running is not zero, $\alpha_T$ at the field value $\phi$ corresponding to direct detection experiments.
We do not reproduce the analytical expressions associated with the calculation of predictions under these models as these are either readily available in Ref.~\cite{Smith:06} or derived through equations outlined there.

\subsection{Direct Detection Observables}

While CMB measurements involve $P_S(k)$ and $P_T(k)$, direct detection measurements involve $P_T(k)$ only.
As we discuss later, the lack of information on the primordial scalar spectrum at direct detection frequencies
complicates simple tests and studies related to slow-roll inflation using data from direct detection experiments alone. 
Furthermore, since the observations involve the relic background today, one must also account for the evolution of the
tensor mode from high redshifts to low redshifts. This evolution of a single gravitational wave mode 
can be written through the massless Klein-Gordon equation
\begin{equation}
\frac{d^2h_k}{d \tau^2}+ 2 \frac{1}{a} \frac{da}{d \tau} \frac{dh_k}{d \tau} + k^2 h_k = 0
\end{equation}
with the boundary conditions $h_k(0) = P_T(k)^{1/2}$ and $\dot{h_k}(0) = 0$.  Here, 
$\tau$ is conformal time. We define $g_k(\tau) \equiv h_k(\tau) a(\tau)$.  With this definition, we are able to rewrite the above equation,
\begin{equation}
\frac{d^2 g_k}{d \tau^2} +\left(k^2 - \frac{1}{a} \frac{d^2 a}{d \tau^2} \right) g_k = 0.
\end{equation}
There are two limiting behaviors for $g_k$: before horizon entry
$g_k \propto a \rightarrow h_k = \rm constant$; after horizon entry $g_k$ will oscillate which implies $h_k$ will oscillate
with an amplitude decreasing as $1/a$ \cite{Turner:93}.  Therefore, the current spectrum of gravitational waves is determined by the primordial
power spectrum and by the rate at which scales enter the horizon (i.e. the evolution of the scale factor): $h = h_0 a_k/a$.  
During radiation domination, $H \propto a^{-2}$,
so that $k = 1/a_k$ and during matter domination $H \propto a^{-3/2}$, so that $k = 1/a_k^{1/2}$.  From these relations,
we find
\begin{eqnarray}
h_k &\propto& k^{-1} \ \rm (Radiation \; Domination) \\
h_k &\propto& k^{-2} \ \rm (Matter \; Domination) \, .
\end{eqnarray}
These scalings are generally quantified in terms of the  transfer function $T_T(k)$ for tensor modes \cite{Turner:93},
such that the tensor power spectrum is related to the one generated during inflation as
\begin{equation}
P_T(k,t=t_0) = P_T(k,t=t_\infty)T^2_T(k) \, .
\end{equation}
We define the ratio of energy density of the gravitational wave background to the critical closure density as
\begin{equation}
\Omega_{\rm GW}(k) = \frac{1}{\rho_c} \frac{d\rho_{\rm GW}}{d \ln k} \, .
\end{equation}
and write
\begin{equation}
\Omega_{\rm GW}(k)h^2=\frac{c^2 k^2 h^2}{6 H_0^2} \langle|h_k|^2 \rangle \equiv A_{\rm GW}P_T(k) \, ,
\end{equation}
where $A_{\rm GW}=2.74 \times 10^{-6}$ at $f=0.1$ Hz \cite{Smith:06} and comes from taking a frequency average of the tensor transfer function  $T_T(k)$.
 The mapping amplitude $A_{\rm GW}$ ignores the
anisotropic stresses in the cosmic fluid, among which free-streaming of relic neutrinos is important \cite{Weinberg:04}. 
The additional damping of tensor modes, however, is only restricted to modes that enter the horizon after neutrino free-streaming or the gravitational
wave background at frequencies below 10$^{-11}$ Hz. Thus, our calculations are not affected by ignoring anisotropic stresses.
We do note that other exotic processes may damp gravitational waves at frequencies of 0.1 Hz \cite{Boyle}, but the physical processes
are mostly speculative and we do not consider them to be important for this calculation.

In addition to the amplitude of the gravitational wave background captured by $\Omega_{\rm GW}h^2$, recent analyses suggest
that it may also be possible to measure the tilt $n_T$ of the gravitational wave background at direct detection frequencies 
\cite{Seto:06}. Thus, we discuss the observables in terms of both  $\Omega_{\rm GW}h^2$ and $n_T$ here.
For direct detection experiments, we consider three possibilities here involving BBO and an improved version of BBO (BBO-grand)
as well as DECIGO. These experiments and their sensitivities are summarized in Ref.~\cite{Seto:06}, and to be consistent with
those calculations, we make use of the same quoted sensitivities here. We do not consider a version of BBO called BBO-lite with
a sensitivity to gravitational waves only down to $\Omega_{\rm GW}h^2$ of $10^{-15}$. As we find later, such an experiment is unlikely to
result in any detection of the gravitational wave background as current limits already rule out a background just above this level.

\begin{figure*}[!t]
\centerline{\psfig{file=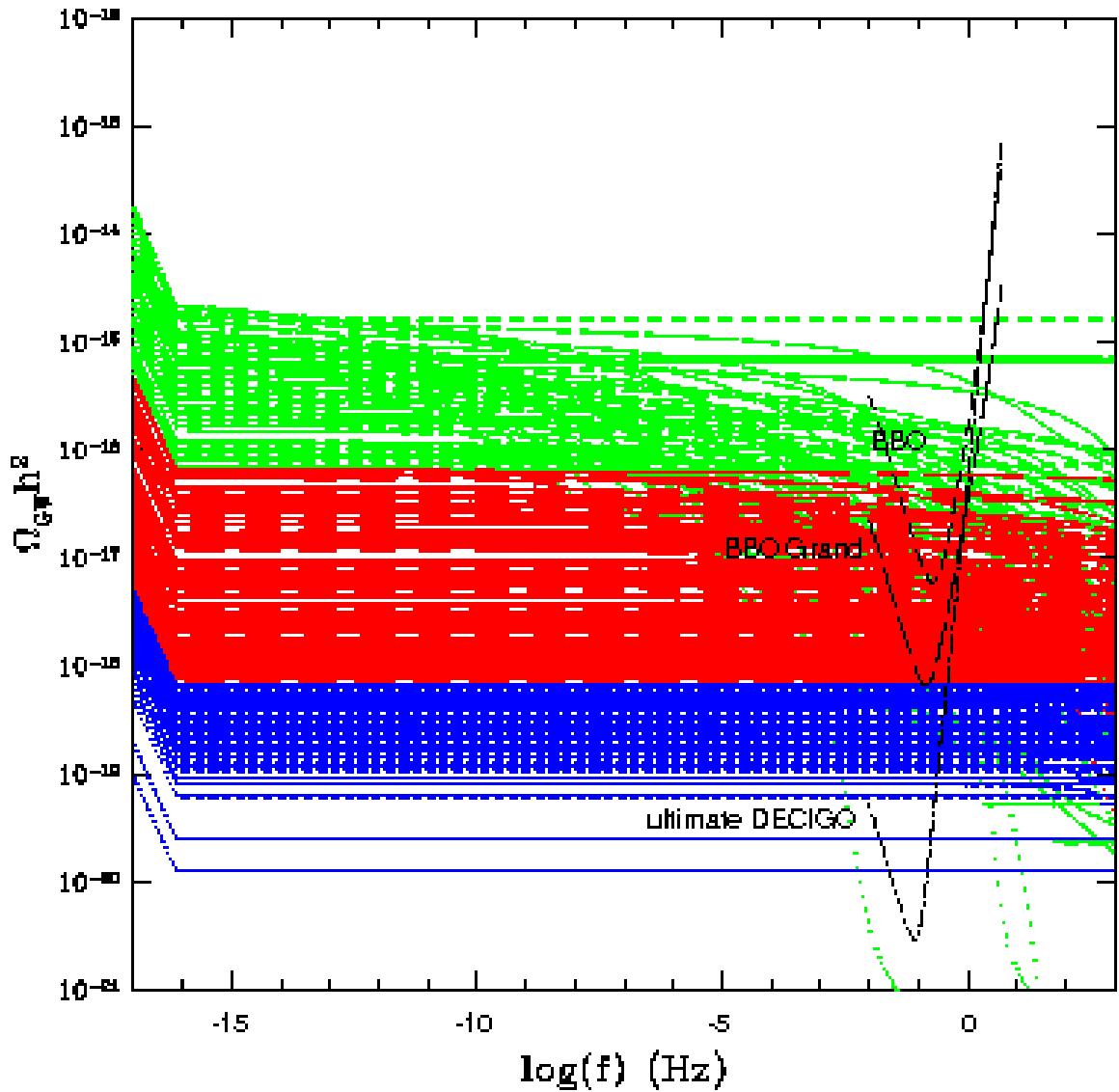,width=6.0in,angle=0}}
\caption{The amplitude of the stochastic gravitational wave background, $\Omega_{\rm GW}h^2$, versus the gravitational wave frequency.
The curves, subdivided by the tensor-to-scalar ratio at CMB scales, represent the expectations based on inflationary models
selected from solving inflationary flow equations and consistent with
WMAP+SDSS data. The models that lead to large tensor-to-scalar ratios
generally have large spectral indices for the gravitational wave background resulting in a damping of the power at frequencies related to
direct detection  experiments. We also show the detector sensitivities for a 5 year-long integration of some of
the options discussed in the literature.  These experiments, from top to bottom, are BBO, an improved version of BBO (BBO-Grand), and an
extremely optimistic version  of the DECIGO (Ultimate DECIGO). 
}
\end{figure*}

\begin{figure}[!t]
\centerline{\psfig{file=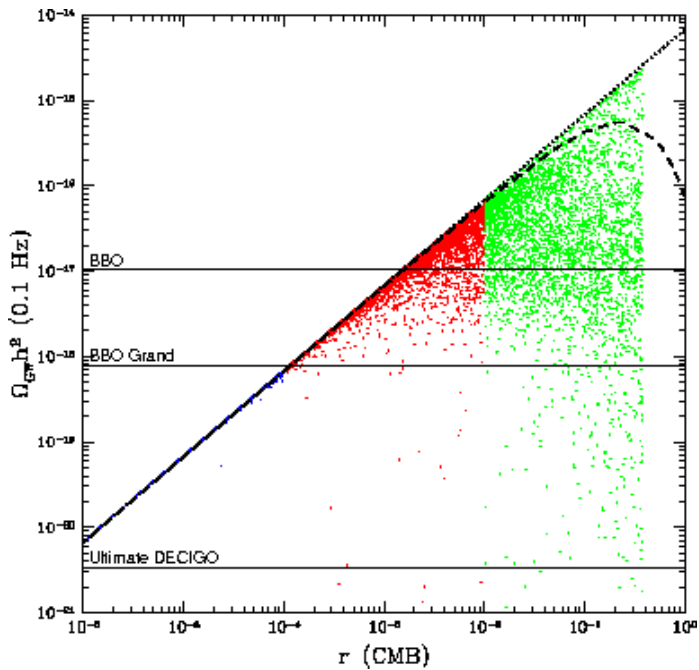,width=3.6in,angle=0}}
\caption{The amplitude  of the gravitational wave background at 0.1 Hz $\Omega_{\rm GW}h^2$ 
plotted in terms of the tensor-to-scalar ratio at CMB scales ($k=0.002$ Mpc$^{-1}$).
For reference, we also show the minimum background amplitude that can be detected with a signal-to-noise ratio of unity with BBO, BBO-Grand, and Ultimate DECIGO.
The relation between $\Omega_{\rm GW}h^2$ and $r_{\rm CMB}$ is such that one can establish an upper limit on the expected amplitude of the gravitational
wave background at a given tensor-to-scalar ratio. The dotted line shows the relation, which we discuss in equation~27.
An experiment such as BBO-Grand can probe slow-roll inflationary models with tensor-to-scalar ratio greater than 10$^{-4}$.
}
\end{figure}

\begin{figure*}[!t]
\centerline{\psfig{file=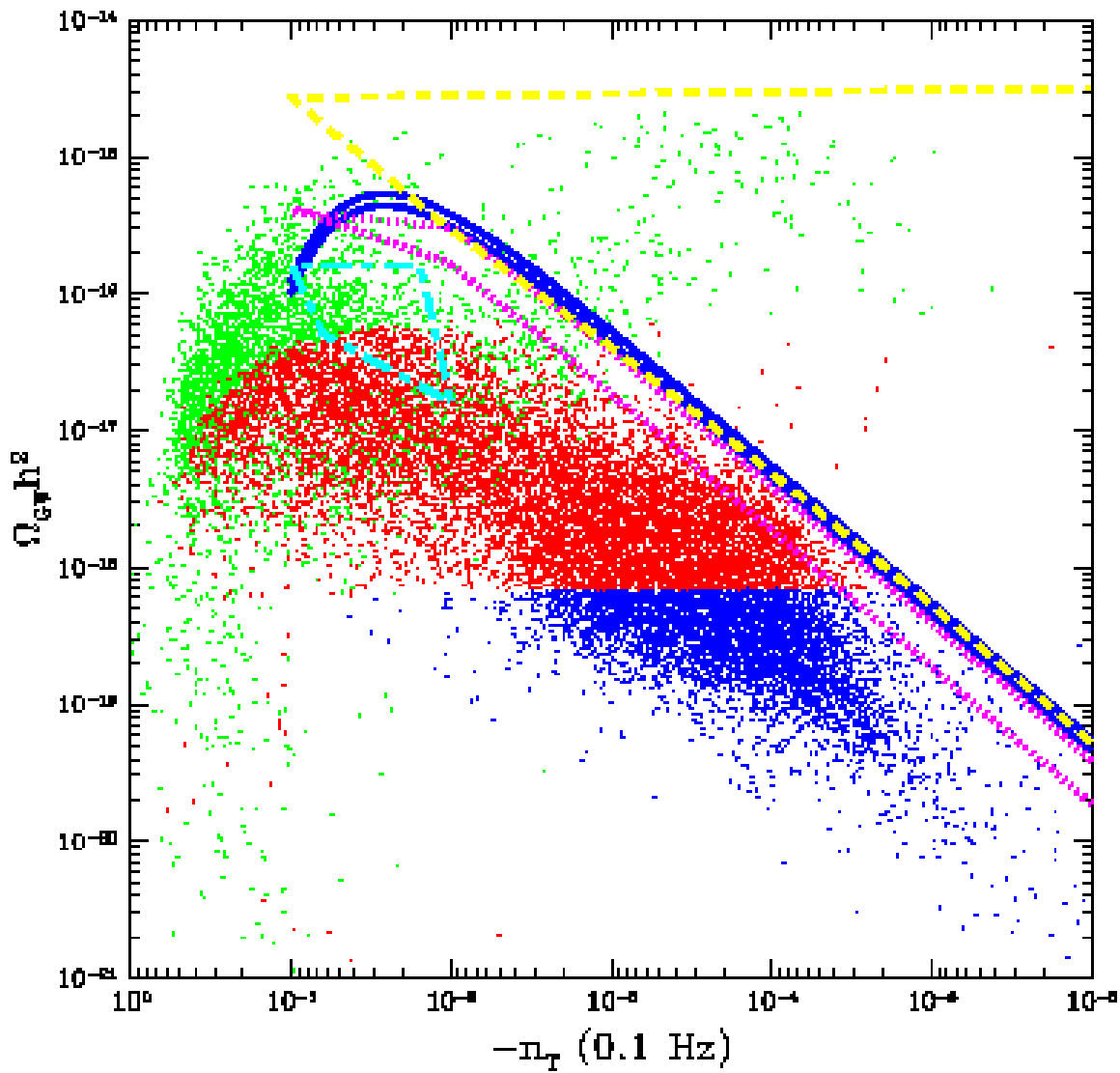,width=6.0in,angle=0}}
\caption{The amplitude of the gravitational wave background $\Omega_{\rm GW}h^2$ at 0.1 Hz 
plotted against the spectral index of the gravitational wave background power spectrum at 0.1 Hz.
We determine this spectral index using the inflationary flow equations with $n_T(\phi)$ determined at the scalar field value $\phi$ corresponding to 0.1 Hz.
In solid, dotted, dot-dashed and dashed curves, we show regimes in the $\Omega_{\rm GW}h^2-n_T$ plane at 0.1 Hz populated by power-law, chaotic, symmetry breaking,
and hybrid models of inflation. In contrast to Figure~1, where there is a large overlap between Monte-Carlo models and analytical models at $k=0.002$ Mpc$^{-1}$,
we find that the predictions based on inflationary flow equations depart significantly from predictions of the analytical models at $k\sim 6.5 \times 10^{13}$.
}
\end{figure*}

\begin{figure*}[!t]
\centerline{\psfig{file=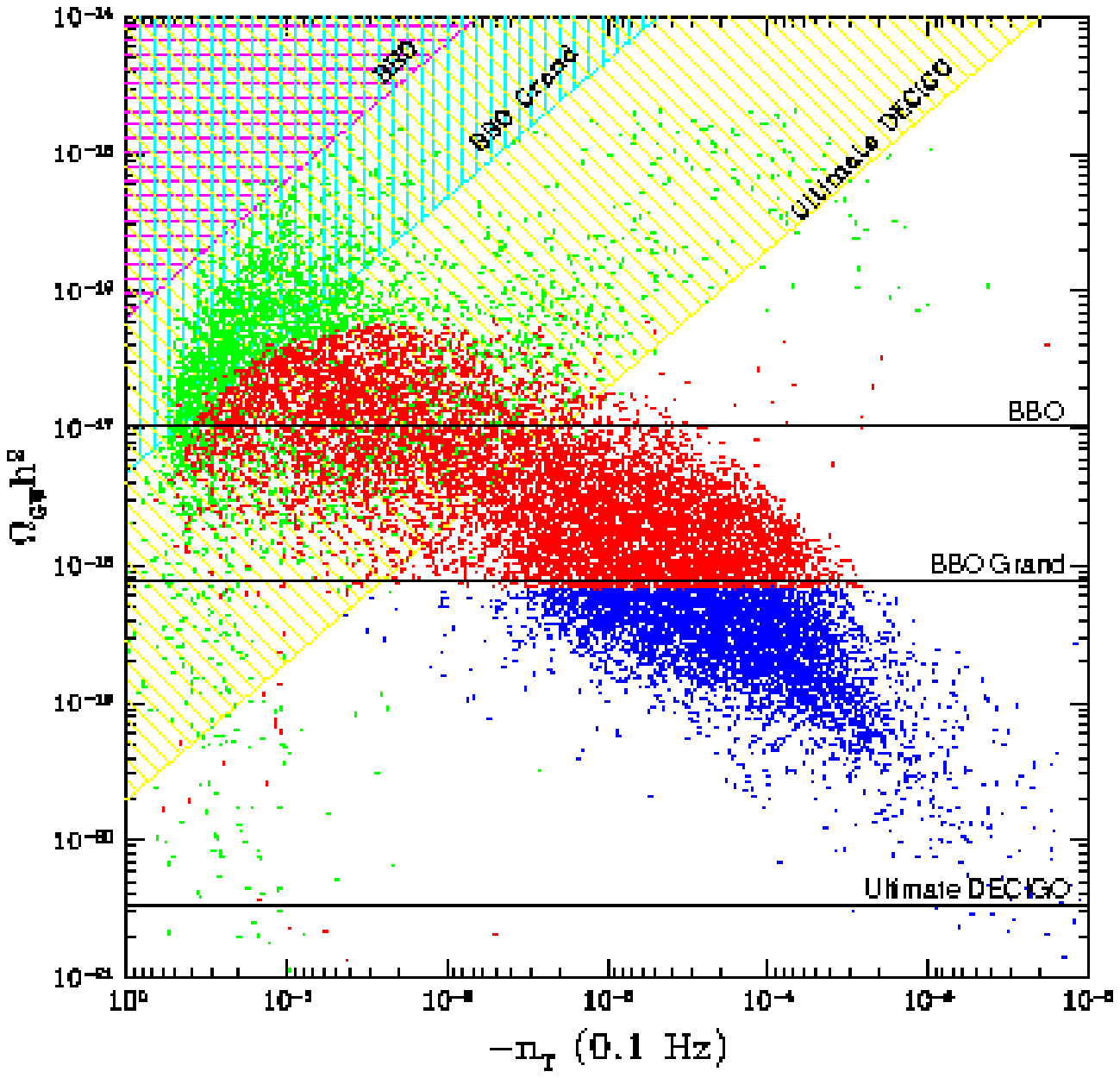,width=6.0in,angle=0}}
\caption{The amplitude of the gravitational wave background at 0.1 Hz plotted in terms of the spectral index of the gravitational wave background power spectrum at 0.1 Hz.
The data points are same as those in the previous figure. Here, we put the measurements in the context of experimental studies.
The horizontal lines show the minimum amplitude of the gravitational wave background detectable with a signal-to-noise ratio of one.
The shaded areas represent the regions in the $\Omega_{\rm GW}h^2-n_T$ plane such
that these detectors can detect both the amplitude and the tensor spectral index $n_T$ at 1$\sigma$ confidence level.
}
\end{figure*}

\section{Results}

\subsection{CMB Predictions and Comparison}

Figure~1 summarizes the plane of tensor-to-scalar ratio versus scalar tilt at CMB scales ($k=0.002$ Mpc$^{-1}$),
 where points represent results of the Monte-Carlo process using the flow equations.
We constrain the parameter space to be $r_{\rm CMB} <0.38$ and $0.92 < n_s < 1.06$ at the 95\% confidence
level following the results of Ref.~\cite{Spergel} and Ref.~\cite{Kinney:06}. 

In addition to lying within this range, we also require that each of the selected models
generates between  46 and 67 e-folds before the end of inflation. The two horizontal lines
represent the expected limits from CMB observations: the published goal of the next generation NASA CMB polarization mission with a tensor-to-scalar
ratio of  $10^{-2}$ \cite{Weiss} and an optimistic limit on the tensor-to-scalar ratio based  on polarized foregrounds (10$^{-4}$). This limit is somewhat
unknown  and estimates suggest that it is between $10^{-3}$ and 10$^{-4}$ \cite{foregrounds}, depending on the frequency
selection, the sky area targeted for CMB observations, and the unknown polarized intensity of dust. Once high
frequency all-sky CMB polarization maps become available with Planck, this foreground-limited tensor-to-scalar ratio will be better established.
To guide our discussion, here, we take the lower limit of 10$^{-4}$.

In addition to inflationary flow equation predictions, we also show the expected range allowed by the model parameter space of several families
of analytical models of inflation. These include the power-law, chaotic, symmetry breaking,  and Hybrid inflation models. The allowed
ranges we highlight here are consistent with previous calculations \cite{Smith:06} and both analytical and Monte-Carlo numerical models are
normalized to the same amplitude of the scalar perturbations at $k=0.002$ Mpc$^{-1}$. As is well known, hybrid inflation models generally allow $n_s > 1$,
while other families generally fill up the parameter space with $n_s < 1$. 

For comparison, in Figure~2, we plot the scalar spectral index against the running of the scalar spectral index, $\alpha_s$, at CMB scales.
The top line indicating an arrow pointed to low values of $\alpha_s$ is roughly the upper limit on the running of scalar tilt with a value
of  0.007 at the 95\% confidence level using WMAP+SDSS data \cite{Kinney:06}.
At the low end, the running is allowed to be as low as -0.13 and all our data points lie above this limit.  While some of our models
lie above $\alpha_s=0.007$, we do not impose a selection based on $\alpha_s$, as we have seen from the literature that the determination of $\alpha_s$ is largely subject to
assumptions coming from model fitting.
As an example, in Ref.~\cite{PeiEas06}, $\alpha_s$ is constrained to be below 0.01 at the 1$\sigma$ confidence level with a prior
of 30 e-folds or more between CMB and the end of inflation.

The corresponding potentials allowed by the inflationary flow analysis in the ($n_s,r_{\rm CMB}$) plane are shown in Figure~3.
The potentials are subdivided based on the tensor-to-scalar ratio at CMB scales. In Figure~3 $\phi=0$ corresponds to
CMB scales while $\phi >0$ represents scales below CMB. The potentials are truncated at the end of inflation when $\epsilon$ reaches a value of unity,
and again, all these potentials lead to between 46 and 67 e-folds of inflation. 
As shown, potentials that give rise to large tensor-to-scalar ratios with $r_{\rm CMB} > 10^{-2}$
cannot be described by simple power-laws or with any particular analytical shape. In fact, almost all of the potentials
cannot be described by simple analytical forms. 

In general, inflationary potentials with $r_{\rm CMB} > 10^{-2}$ have $\Delta \phi > \mpl$, where $\Delta \phi$ corresponds to the number of e-folds 
from CMB horizon exit to end of inflation. This result is well known in terms of the
 Lyth-bound \cite{EfsMac05}.  In Fig.~4, we summarize the relation between the number of e-folds $N$ and the field value $\phi$ for potentials that give
rise to $r_{\rm CMB} > 10^{-2}$ (left panel) and  $r_{\rm CMB} < 10^{-4}$ (right panel). As can be seen in Fig.~3, potentials that lead to large tensor-to-scalar
ratio at CMB scales have structure such that  towards the end of inflation potentials begin to flatten followed by a drop in the amplitude
as $\epsilon \rightarrow 1$.  Since the number of e-folds between two field values can be written as 
\begin{equation}
N(\Delta \phi) \approx  \frac{8\pi}{\mpl^2} \int_{\phi_i}^{\phi_f} \frac{V(\phi)}{V'(\phi)} d\phi \, ,
\end{equation}
and since $V'(\phi)$ becomes smaller towards the end of inflation, one finds a large change in the number of e-folds. In Fig.~4 we show some
example relations between $N$ and $\phi$ which highlight the fact that most potentials that lead to $r_{\rm CMB}>10^{-2}$ behave such that the last 20 to 30 e-folds
of inflation happen suddenly over a small duration of $\Delta \phi$. The models with steep slopes around $\phi=2$ are prime examples of this behavior.
 Similar potential shapes have been noted before and explained in terms of the relation between $H(N)$ and $N$ in Ref.\cite{Efstathiou}.
As discussed there, direct detection experiments that are only sensitive to gravitational wave backgrounds with large $r_{\rm CMB}$  values
will be probing potentials with this abrupt behavior. None of the analytical models have behaviors like this. In addition to the changes in the number
of e-folds, the shapes of these potentials are such that, as we discuss in the next subsection, the tensor tilt is strongly scale dependent.
Because of the negative value of the scale dependent tensor tilt, the gravitational wave background
 amplitude at direct detection scales becomes lower than the case where tensor tilt is scale independent.

\subsection{Predictions for Gravitational Wave Observations}

With observables at CMB scale defined and models selected based on the WMAP+SDSS constraints, we now extend the discussion to consider
the predictions for gravitational wave detection at direct detection frequencies. Figure~5 shows the prediction for
$\Omega_{\rm GW}h^2$ as a function of frequency, where in addition to the power spectrum of tensor perturbations generated
during inflation, we also take into account the evolution through the transfer function (See, Section IIC).
As can be seen in Figure~5, the potentials that have large variations lead to significant spectral indices for the tensor
power spectrum, while for $r_{\rm CMB} < 10^{-4}$ the tensor tilt is insignificant and the approximation that $n_T \approx 0$ is reasonably accurate.
The correspondence between large $r_{\rm CMB}$ and large $n_T$ impacts planning of direct detection experiments.
For reference, in Figure~5, we also plot the sensitivity curves of three of the experimental options that are routinely
discussed in the literature \cite{Smith:06,Smith,Efstathiou,Coo:05}. The sensitivity curves are the same ones shown in
Ref.~\cite{Seto:06} which were used to estimate how well these interferometers can study the gravitational wave background amplitude and the spectral index.

Figure~6 shows the relation between the amplitude of the gravitational wave background at a frequency of 0.1 Hz and the tensor-to-scalar
ratio at CMB ($k=0.002$ Mpc$^{-1}$). Such a relation has been previously discussed in Refs.~\cite{Smith,Efstathiou}.
The data points are distributed such that one clearly notices an upper limit on $\Omega_{\rm GW}h^2$
at a fixed $r$, which we determine to be
\begin{equation}
\Omega_{\rm GW}h^2 < 6.72 \times 10^{-15} r  \left(\frac{P_s(k_{\rm CMB})}{2.45 \times 10^{-9}}\right)\left(\frac{A_{\rm GW}}{2.74\times 10^{-6}}\right) \, .
\end{equation}
This limit agrees with the result of Ref.~\cite{Efstathiou} when converting our scalar power spectrum normalization and the transfer function to their
values. This upper limit comes from the fact that under slow-roll inflation, $n_T$ has a negative value if it is not zero.
 In Fig.~6, for comparison, we also produce the prediction under the power-law potential for inflation. At low values of the tensor-to-scalar
ratio at CMB, there is a tight correlation between $\Omega_{\rm GW}h^2$ and $r_{\rm CMB}$, which falls along the power-law expectation.

With $r_{\rm CMB}<0.38$ at the 95\% confidence level from current observations, the derived upper limit on the gravitational wave background at 0.1 Hz,
under standard slow-roll inflation, is such that $\Omega_{\rm GW}h^2 < 2.55 \times 10^{-15}$. Experimental options, such as the BBO-lite 
\cite{Seto:06},  with a limiting sensitivity of 10$^{-15}$ in $\Omega_{\rm GW}h^2$, are clearly unfavored as they will only explore an extremely
small parameter space. Moreover, CMB experiments such as {\it Planck} will soon explore the tensor-to-scalar ratio down to a limit of 0.1,
while the next generation CMB polarization mission targets to improve this down to a limit of 0.01. In terms of the amplitude alone,
a direct detection experiment must have sensitivity below $6 \times 10^{-17}$ in $\Omega_{\rm GW}h^2$ to be both useful in terms of
model selection and studies related to inflation. Furthermore, even if CMB experiments were to detect a primordial tensor component with
a tensor-to-scalar ratio above 0.01, an experimental option such as the standard-BBO \cite{Seto:06} is not preferred as 
many of our inflationary models are such that with increasing $r_{\rm CMB}$, $n_T$ (which is negative) also increases in magnitude.
This results in a large suppression of power at direct detection frequencies as shown in Fig.~5 for $\Omega_{\rm GW}h^2$ versus frequency for 
curves with $r > 0.01$.

The preference to select a direct detection experiment with adequate sensitivity down to a CMB tensor-to-scalar ratio of 10$^{-4}$ is simply a criteria based on
 slow-roll inflationary models.
It could be that, under non-standard hypotheses, one can generate blue-tilted models for tensor fluctuations
such that one generates a large gravitational wave background at frequencies around 0.1 Hz.  We refer the reader to Ref.~\cite{Efstathiou}
for a discussion of such possibilities. We, however, find less motivation to design a target for 
a direct detection experiment based on non-standard criteria alone since existing cosmological data rules out most of these cases.

\begin{figure*}[!t]
\centerline{\psfig{file=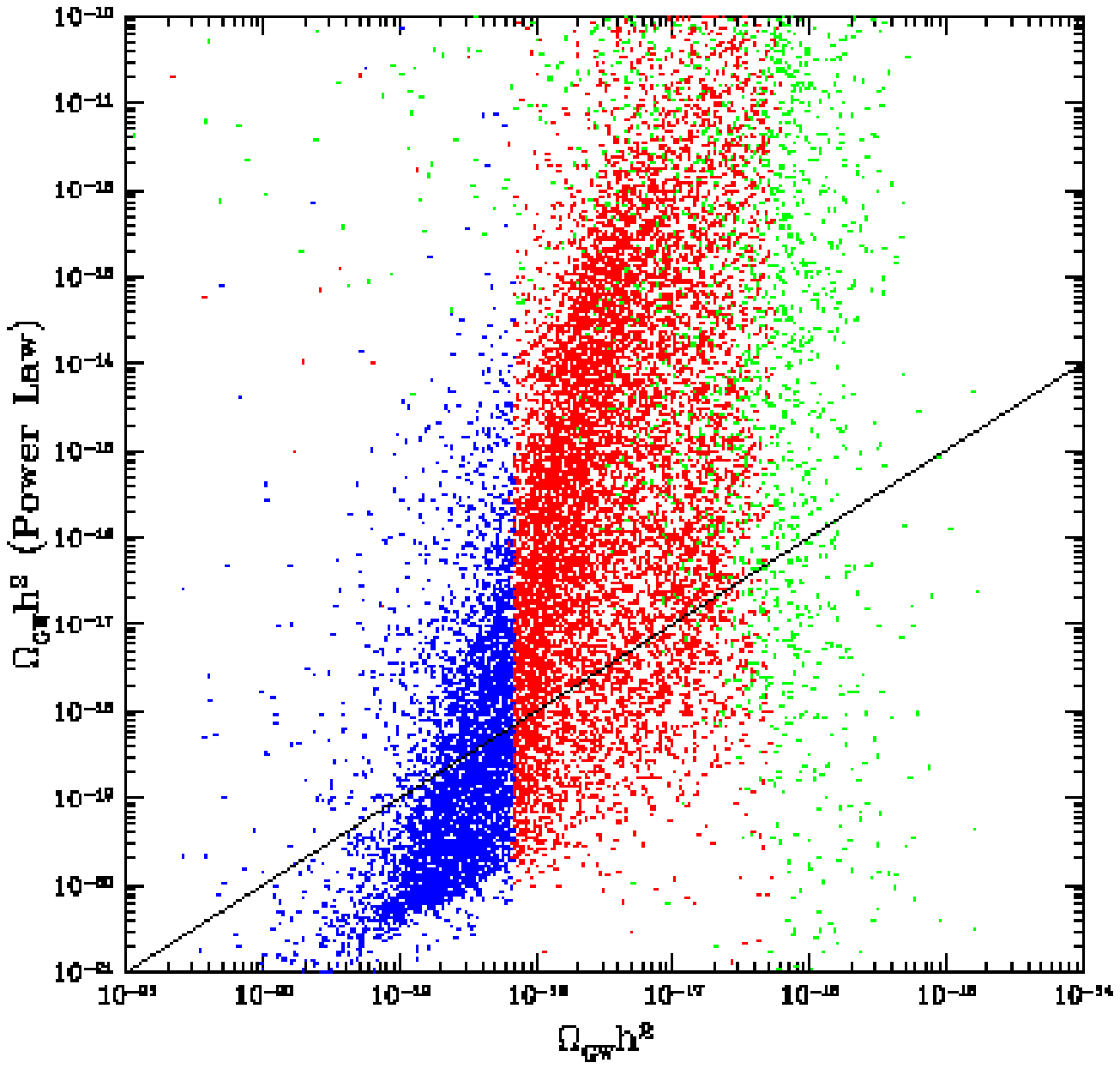,width=3.6in,angle=0}
\psfig{file=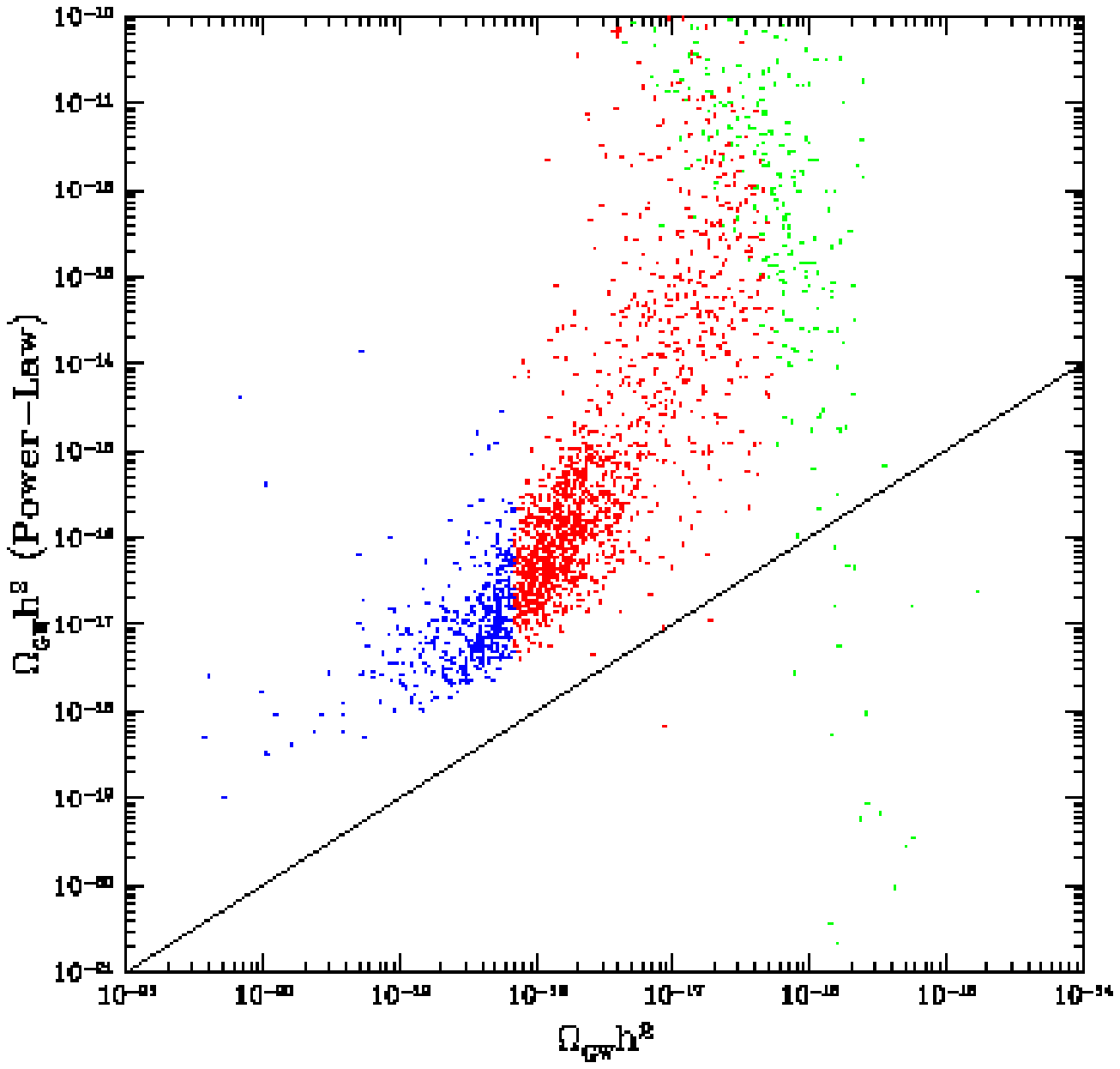,width=3.6in,angle=0}}
\caption{$\Omega_{\rm GW}h^2$ at 0.1 Hz plotted against the estimated $\Omega_{\rm GW}h^2$ assuming a power-law expansion.
The solid line at an angle represents the case where $\Omega_{\rm GW}h^2=[\Omega_{\rm GW}h^2]_{\rm power-law}$.
The left panel shows the relation for all data points, while the right panel selects points with $0.99 < n_s < 1.01$ at CMB scales.
The power-law approximation is based on an extrapolation of CMB observations using the values $r_{\rm CMB}, n_S(k_{\rm CMB}), \alpha_S(k_{\rm CMB})$,
and the assumption that $n_T=-r_{\rm CMB}/8$ and $\alpha_T=0$ at all scales. This fails to capture the amplitude 
of the background predicted in our exact numerical calculation. The power-law approximation
leads to a general overestimate of the gravitational-wave  background at direct detection frequencies. While some of the scatter in the
left panel is due to the large range of $n_s$, by selecting inflationary models with a small range in $n_s$, we show that the
overestimate is not restricted to either large or small $n_s$ models at CMB scales.}
\end{figure*}

\begin{figure*}[!t]
\centerline{\psfig{file=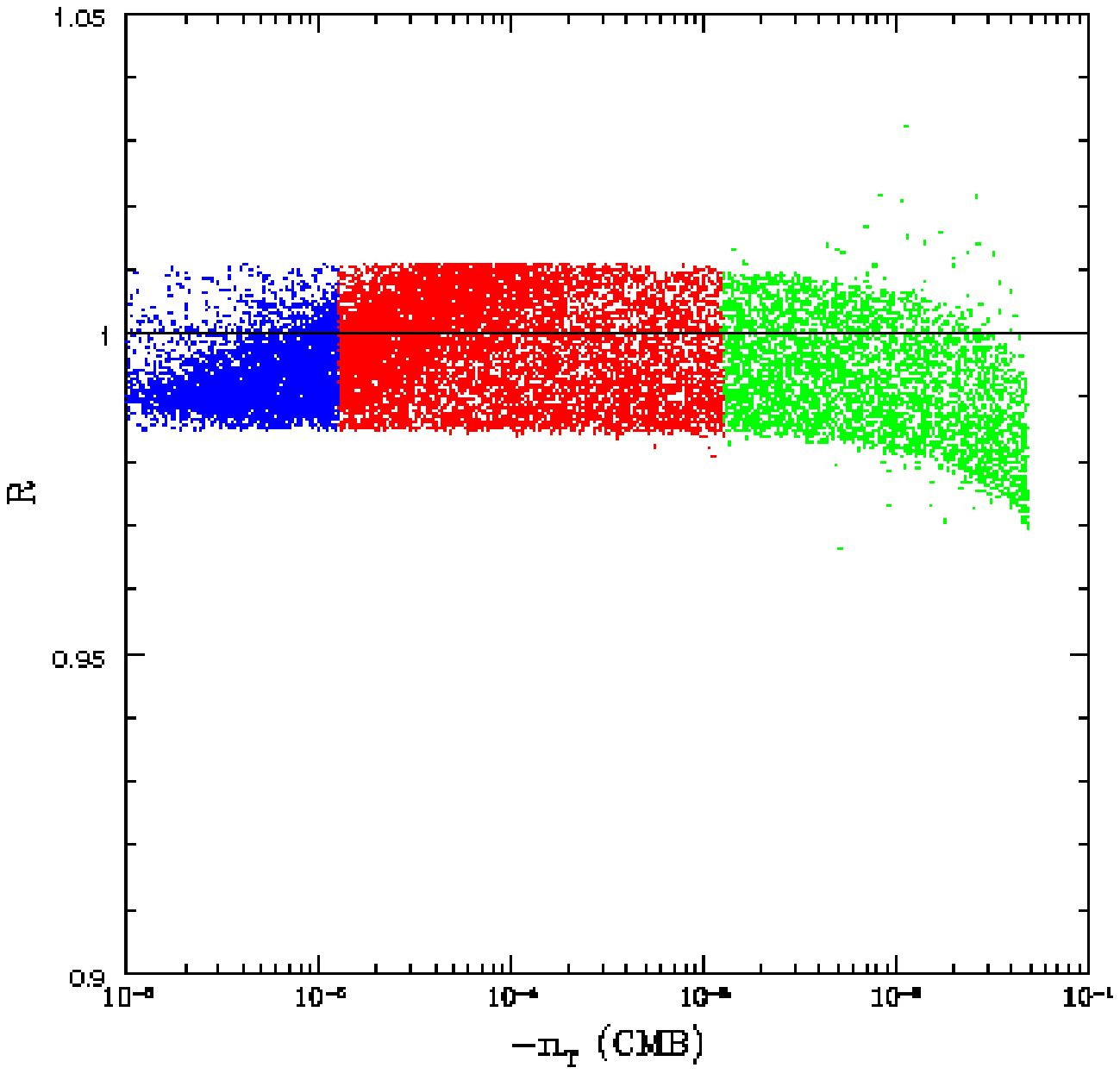,width=3.6in,angle=0}
\psfig{file=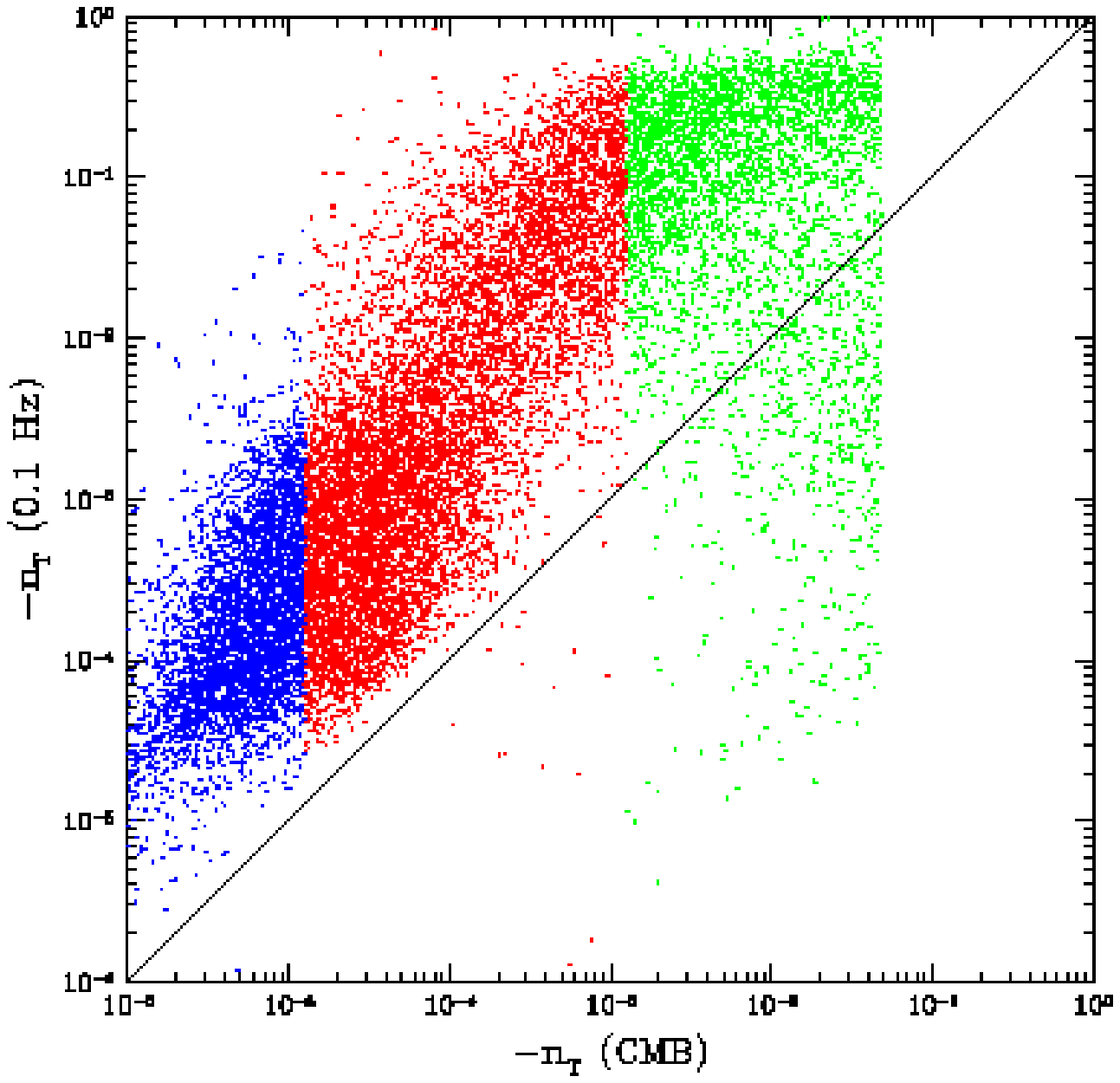,width=3.6in,angle=0}}
\caption{
{\it Left}: The consistency relation between the tensor amplitude $r$ and the tensor tilt $n_T$. Under first order slow roll
expansion, $r=-8n_T$. Here, we plot $R \equiv r/(-8n_T)$ as a function of $n_T$ for measurements at CMB scales.
This consistency relation is challenging to measure with CMB data alone due to lack of information in the CMB power spectrum for a reliable
measurement of $n_T$. The spread in R and the tendency for R to dip slightly below one for large $n_t$ is a result of the second order terms
used to calculate $r$ and $n_T$. For $-n_T > 0.01$, one sees a departure from $R=1$ due to the fact that for these models,
$n_T > -2 \epsilon$ and  the $\epsilon^2$ term is more significant for $n_T$ than for $r$.
{\it Right:} In a recent study, Smith et al., it was suggested that the inflationary consistency relation can be tested by
combining CMB observations with direct detection measurements under the assumption that $n_T(k_{\rm CMB}) \approx n_T({\rm 0.1 Hz})$.
Here, we plot the 
tensor spectral index at 0.1 Hz versus the same spectral index at CMB showing that $n_T$ at 0.1 Hz is larger than the value at CMB scales
and the assumption that $n_T(k_{\rm CMB}) \approx n_T({\rm 0.1 Hz})$ is not consistent with the models predicted under slow-roll inflationary flow
equations. In general, using direct detection slope to approximate tensor tilt at CMB scales results in a bias of an order of
magnitude or more.
We explain this difference in terms of the scale dependence of the tensor tilt and higher order corrections such as $\alpha_T$.}
\end{figure*}

\begin{figure}[!t]
\centerline{\psfig{file=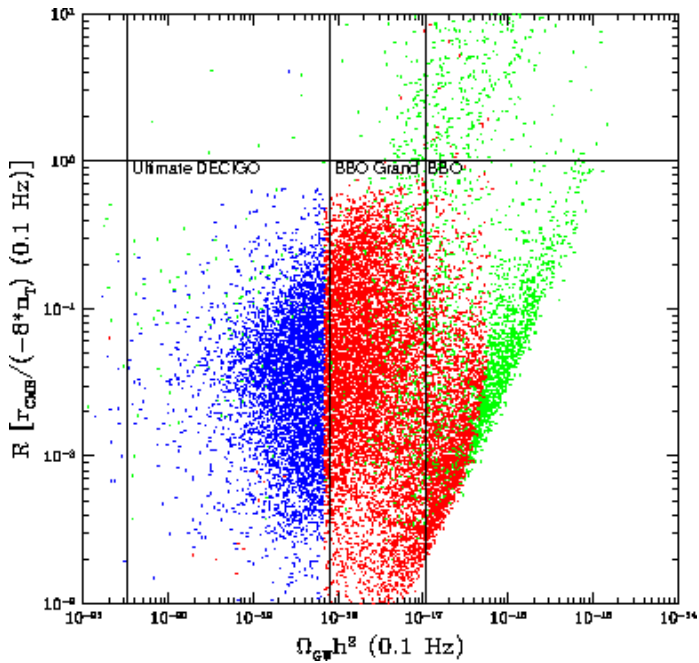,width=3.6in,angle=0}}
\caption{
The consistency relation between the tensor amplitude $r$ and the tensor tilt $n_T$, $R\equiv -r/8n_T$
as a function of the GW background amplitude $\Omega_{\rm GW}h^2$ at 0.1 Hz under the assumption that $n_T(k_{\rm CMB}) \approx n_T({\rm 0.1 Hz})$ and  
using $r_{\rm CMB}$ to estimate $R$.
Since $n_T({\rm 0.1 Hz}) > n_T(k_{\rm CMB})$ by at least an order of magnitude, $R$ departs from the expected value of one by a factor of 10 or more.
The figure demonstrates that one cannot naively combine CMB tensor-to-scalar ratio with the tensor tilt
at 0.1 Hz to test the consistency relation. The vertical lines mark the limits related to $\Omega_{\rm GW}h^2$ for three of the experiments while  
the horizontal line $R=1$ marks the expectation under first-order slow-roll inflation.}
\end{figure}

\begin{figure*}[!t]
\centerline{
\psfig{file=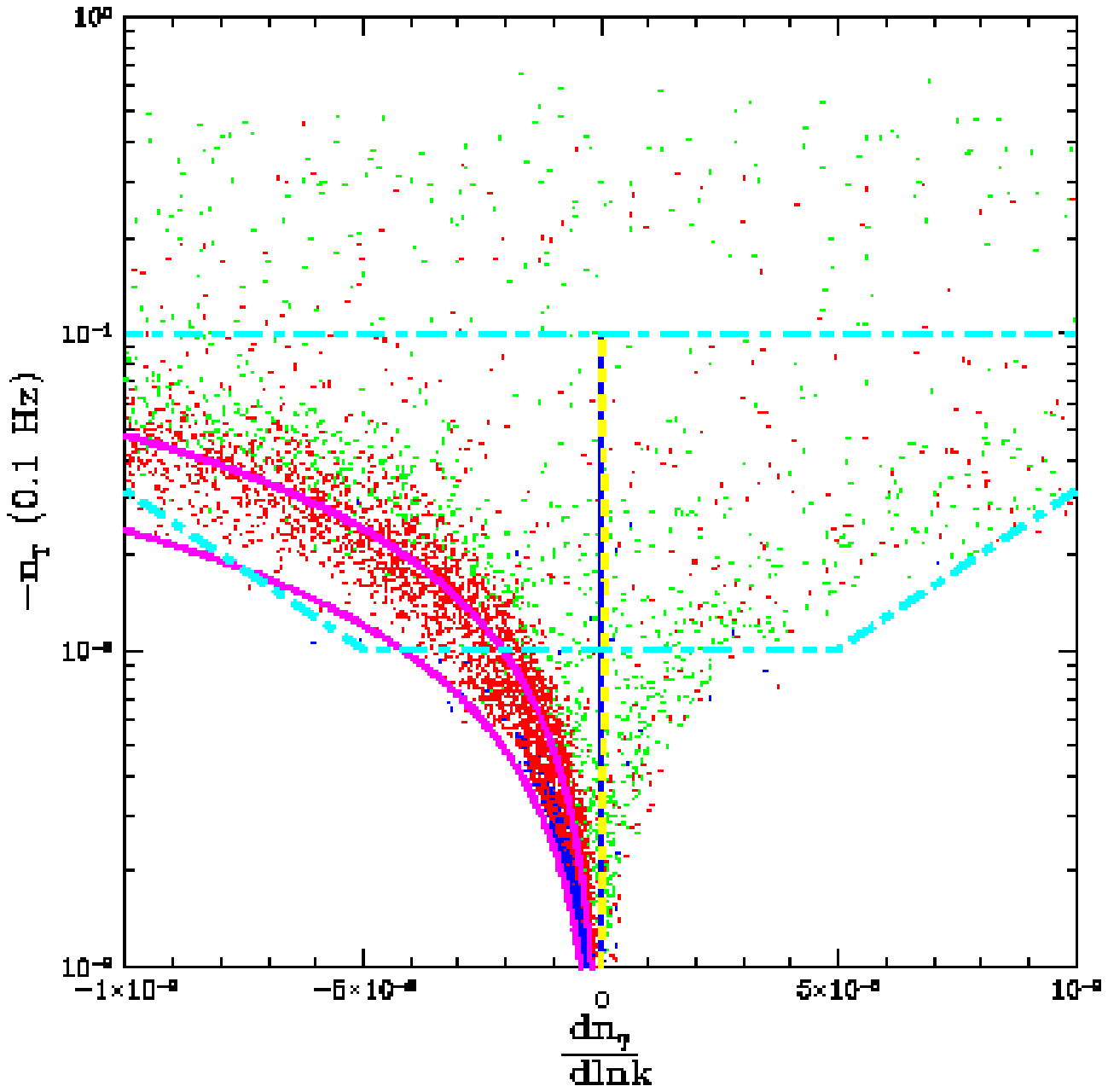,width=3.6in,angle=0}
\psfig{file=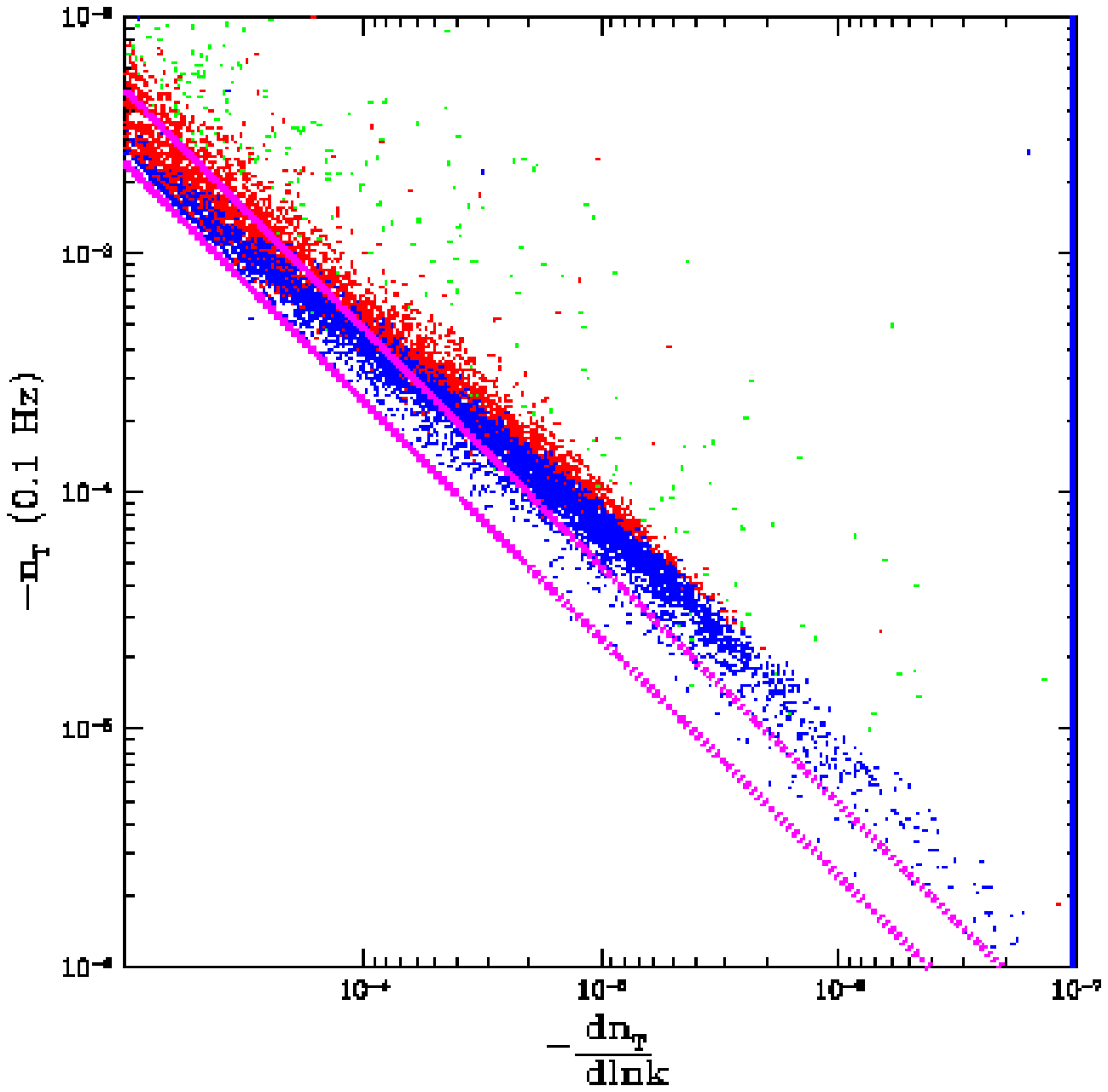,width=3.6in,angle=0}}
\caption{The tensor spectral index $n_T$ versus the running of the tensor tilt, $\alpha_T$ at 0.1 Hz. The data points show the 
inflationary flow predictions, while the
shaded regions show the expectations under analytical model with solid, dotted, dot-dashed, and dashed lines showing power-law, chaotic, symmetry breaking, and
hybrid inflation models, respectively. In certain analytical models, such as power-law, $\alpha_T=0$ over all scales. While the left panel
shows $\alpha_T$ in a linear scale, the right panel concentrates on the low $\alpha_T$ region with values around 10$^{-5}$.
}
\end{figure*}

\subsection{Experimental Detection}

Figures~7 and 8 highlight the gravitational wave amplitude, in $\Omega_{\rm GW}h^2$ as a function of the tensor tilt
at a frequency of 0.1 Hz. In addition to the Monte-Carlo models generated from the
inflationary flow equations, we also show the model predictions obtained from analytical models.
These calculations reproduce results in Ref.~\cite{Smith:06}. Figure~8 plots the same information as in Figure~7, except that
in this plot we have put the model distribution in the context of observational measurements. The horizontal lines in
Figure~8 show the amplitude of the gravitational wave background that lead to
a signal-to-noise ratio (SNR) of one with each of these experiments. In terms of the amplitude,
BBO, BBO-grand and  Ultimate DECIGO are expected to detect gravitational wave
backgrounds with amplitudes $\Omega_{\rm GW}h^2=X \times 10^{-18}$ 
where $X=10,1,0.01$ respectively, consistent with  previous discussions regarding these experiments \cite{Seto:06,Smith}.

In addition to the amplitude of the gravitational wave background, it is also useful to consider
the possibility that these detectors can be used for a reliable measurement of the tensor spectral index.
Following the Fisher matrix calculations in Ref.~\cite{Seto:06}, we determine that for a one-sigma detection of the tensor
spectral index in addition to the gravitational wave background amplitude, it is required that 
one roughly satisfies the relation $|n_t| (\Omega_{\rm GW}h^2/6\times 10^{-18})=X$.
As shown by the shaded regions in Figure~8, the requirement that laser interferometers detect both the amplitude of the
gravitational wave background and the tensor spectral index leads to a significant reduction in the model
parameter space probed by these detectors. Models that have both large $\Omega_{\rm GW}h^2$ and $n_T$ may be
studied reliably. But even experiments such as DECIGO will not be able to detect values of $|n_T|$ less than $10^{-3}$ despite their ability
to detect the gravitational wave amplitudes for nearly all Monte-Carlo models.

\subsection{Testing Inflation: CMB vs. Direct Detection}

While the above discussion considered the extent to which the gravitational wave background can be measured with planned interferometers such
as BBO and DECIGO, we now address how CMB information can be put to use in determining the prospects for 
direct detection measurements. This is motivated by the fact that ongoing and upcoming CMB polarization measurements will
improve the existing constraints from WMAP on the CMB tensor-to-scalar ratio and the scalar tilt and running of
the scalar tilt. With predictions from CMB data, one can also optimize the sensitivity requirements 
for a detection of the gravitational wave background. Here, we address the extent to which CMB observations can be
extended to study the gravitational wave background at a scale 16 orders of magnitude larger than CMB scale.

As a first approximation,  one can make use of a simple power-law extrapolation 
to establish the expected gravitational wave background at direct detection frequencies \cite{Ungarelli:05}. To test the accuracy of such
an extrapolation, we compare the predictions using exact numerical flow equations with the predictions calculated using a simple 
power-law extension between CMB and direct detection scales. For this comparison, we first write approximate forms for the scalar and tensor power spectra as
\begin{eqnarray}
P_S(k)& \approx&P_S(k_{\rm CMB}) \nonumber \\
&& \quad \times \left(\frac{k}{k_{\rm CMB}}\right)^{n_S(k_{\rm CMB})-1+\frac{1}{2}\alpha_S(k_{\rm CMB}) \ln\left(\frac{k}{k_{\rm CMB}}\right)} \nonumber \\
P_T(k)& \approx&P_T(k_{\rm CMB})\left(\frac{k}{k_{\rm CMB}}\right)^{n_T(k_{\rm CMB})-1+\frac{1}{2}\alpha_T(k_{\rm CMB})} \, .
\end{eqnarray}
We can write the gravitational wave background at direct detection frequencies as 
\begin{eqnarray}
\Omega_{\rm GW}h^2 &=& A_{\rm GW}P_T({\rm k=0.1 Hz}) \nonumber \\
P_T({\rm k=0.1 Hz})&=&r_{\rm 0.1 Hz} P_S({\rm k=0.1 Hz}) \, .
\end{eqnarray}
We assume that the tensor background satisfies the slow-roll
consistency relation of  $n_T=-r/8$ and that the tensor power spectrum is not running with $\alpha_T=0$. This allows us to write an approximate equation for
the tensor-to-scalar ratio at direct detection scales as
\begin{eqnarray}
&&r_{\rm 0.1 Hz} \approx r_{\rm CMB} \\
&&\quad \times \left(\frac{k}{k_{\rm CMB}}\right)^{n_S(k_{\rm CMB})-1+\frac{1}{2}\alpha_S(k_{\rm CMB}) \ln\left(\frac{k}{k_{\rm CMB}}\right) + r(k_{\rm CMB})/8} \, .
\end{eqnarray}
Using this relation and the approximate scalar power-spectrum with $n_S(k_{\rm CMB})$ and $\alpha_S(k_{\rm CMB})$ in equation~(28), 
one can approximate the gravitational wave background amplitude, 
$[\Omega_{\rm GW}h^2]_{\rm power-law}$ at 0.1 Hz. In Fig.~9, we show the comparison between
$[\Omega_{\rm GW}h^2]_{\rm power-law}$  and the exact amplitude of the gravitational wave background, $\Omega_{\rm GW}h^2$, 
as calculated through numerical solutions to the flow equations.
Relative to the exact $\Omega_{\rm GW}h^2$  values that are generally below 10$^{-15}$, $[\Omega_{\rm GW}h^2]_{\rm power-law}$ 
ranges as high as $10^{-10}$ and over-predicts the gravitational wave background at direct detection frequencies.
The differences are mostly significant for models that involve large CMB tensor-to-scalar ratios. 
These  models have a small, but non-negligible,  scale dependence in the tensor tilt.

Note that $n_T=-r/8$ and $\alpha_T=0$ are simple assumptions that do not hold for general inflationary models, except in the case of power-law inflation \cite{Smith:06}.
These two assumptions also hold well for hybrid models. They do not hold, however, for chaotic and symmetry-breaking potentials as they
often produce significant tensor running (see Figure~12). The predictions under the Monte-Carlo slow-roll inflation models depart
significantly from analytical models where $\alpha_T$ is mostly zero at direct detection scales. This departure is also related to the reason why
the gravitational wave background amplitude cannot be predicted by extrapolating CMB information to a scale roughly 16 orders of magnitude higher as shown in Figure~9.
 As highlighted in Figure~7, with non-zero $\alpha_T$, chaotic and symmetry-breaking models come closest in approximating the
slow-roll numerical predictions at 0.1 Hz, though the agreement is still small. Compared to all these analytical models,
numerical models lead to a substantial scale dependence in the tensor tilt.

This disagreement is a concern for a simple comparison between CMB observations and direct detection experiments.
 At CMB scales, the observations are limited to
$r_{\rm CMB},n_S(k_{\rm CMB})$ and $\alpha_S(k_{\rm CMB})$.  A direct
detection experiment will measure $\Omega_{\rm GW}h^2$ and $n_T({\rm 0.1 Hz})$.
This limited information cannot establish a unique inflationary potential.
 Suggestions have been made that the amplitude of tensor components can be used to establish the slope of the inflaton
potential $V(\phi)$ between CMB and direct detection scales. While this is, in principle, possible, the constraints will not
be unique unless  a particular class of analytical inflationary models  is advocated. Under the general scenario,
where $V(\phi)$ can vary such that the spectral indices are scale dependent, one cannot simply connect CMB scales with direct detection frequencies
through a power-law extrapolation. 

This complication arises due to the large difference in the two scales. This difference also complicates simple
 tests on slow-roll inflation. As an example, we revisit the proposed test in Smith et al. \cite{Smith} on the consistency relation of
$r=-8n_T$, where they assumed $n_T(k_{\rm CMB})\approx n_T(k={\rm 0.1 Hz})$. They made this assumption because 
CMB observations are unlikely to measure the tensor tilt with adequate accuracy due to confusion with lensing B-modes, 
while direct detection experiments, especially with adequate sensitivity, can make a measurement of $n_T$. Using the  assumption that
$n_T(k_{\rm CMB})\approx n_T(k={\rm 0.1 Hz})$ combined with 
the tensor-to-scalar ratio at CMB scales, Smith et al. advocated a way to test the consistency relation of the form
$r_{\rm CMB}=-8n_T (k={\rm 0.1 Hz})$, though resulting constraints
for experiments under consideration were not that impressive. For example, the combination of Ultimate-DECIGO and
{\it Inflation Probe}, at best, would constrain this relation to a level of 20\% if the tensor-to-scalar ratio is between 0.01 and 0.1.
The test of the consistency relation is affected by the fact that, relative to the fractional error on $r$ at any wavenumber, one must measure $n_T$ with
a fractional error that is 8 times better.

Beyond these measurement issues, there are also fundamental problems from the theory side. To study the extent to which the assumption 
that $n_T(k_{\rm CMB})\approx n_T(k={\rm 0.1 Hz})$  is accurate, in Fig.~10 (right panel), we plot $n_T(k={\rm 0.1 Hz})$ versus $n_T(k_{\rm CMB})$.
As shown, there is a large difference between the two tilts such that $|n_T(k={\rm 0.1 Hz})| > |n_T(k_{\rm CMB})|$. The consistency
relation will be violated if one blindly substitutes $n_T(k_{\rm CMB})\approx n_T(k={\rm 0.1 Hz})$ (see, Figure~11), though as shown in the left-panel of
Figure~10, our predictions are accurate to 1\% level of the consistency relation at CMB with $r_{\rm CMB}$ and $n_T(k_{\rm CMB})$.

To show one aspect from which the complications arise, in Fig.~12, we plot the tensor tilt at 0.1 Hz
as a function of the running of the tensor tilt $\alpha_T$, again at 0.1 Hz. For comparison, we also plot model predictions for the running of the
tilt through analytical models, where some models, such as the power-law potential, are such that $\alpha_T=0$ exactly. This large running
is a reflection that the tensor tilt is largely scale dependent. Simply using the tilt
at one wavenumber for the tilt at another wavenumber will lead to incorrect results. Though the running is small,
the replacement of tilt at either CMB or direct detection scales with the value of the other  is problematic since
the two scales are largely different. While previous studies hailed the large difference in wavenumber between CMB and direct detection
experiments as a ``blessing'' to further study inflation, for arbitrary shaped $V(\phi)$ models that are generated numerically,
we find that the situation is too complex for the simple comparison.
The large difference in scale requires a careful understanding of how parameters behave in between the scales, since the potentials are complicated
and do not follow a power-law behavior of $\phi$. And since observations are limited to two scales, making a potential reconstruction is difficult.
This, however, mostly applies to the case where one is comparing observations to establish information related to an arbitrary $V(\phi)$. If one assumes
a prior shape for $V(\phi)$, such as through one of the analytical models, then the behavior can be predicted analytically and the two scales
can be matched to the extent that the data allow.

Instead of combining CMB and direct detection information to test the slow-roll consistency relation, 
perhaps, it is best to improve the measurement of $n_T(k_{\rm CMB})$ with CMB
polarization data, by improving the sensitivities so that confusing lensing B-modes are separated from primordial B-modes \cite{HuOka02},
and then using CMB data alone to study the consistency relation. On the other hand, with the feasibility of a direct detection,
it is also useful to investigate if there is way to use information at direct detection frequencies for a test of inflation.
Unfortunately, due to lack of information on the scalar or density perturbations at $k \sim 10^{13}$ Mpc$^{-1}$
it is unlikely that information on the tensor power spectrum at direct detection frequencies alone will be useful to 
establish slow-roll inflation.  On the other hand, the complications in connecting observations to an underlying model
we have outlined above may start to improve if information on scalar perturbations becomes available at a wavenumber different from that of CMB
but between CMB and direct detection scales. Then there would be additional data points to interpolate between the two largely separated scales. In this respect,
in an upcoming paper, we will return to such a study by concentrating on the power-spectrum of 21-cm
background at $z \sim 50$ to 100 that provides information on the primordial power spectrum out to $k \sim$ a few tens Mpc$^{-1}$.

\section{Summary}

In addition to density perturbations, inflationary models of the early universe generally predict a stochastic background of
gravitational waves or tensor fluctuations. By making use of the inflationary flow approach for single field models and 
with predictions normalized to results from cosmic microwave
background and large scale structure data,
 we discuss the expected properties of the gravitational wave background from inflation
at scales corresponding to direct detection experiments with laser interferometers in space.
We complement the Monte-Carlo calculations based on the inflationary flow equations by including predictions expected under
 several classes of analytical inflationary models, including models involving hybrid inflation.
We find that an improved version of {\it Big Bang Observer} (BBO-grand)
can be used to detect a  gravitational wave background corresponding to a
tensor-to-scalar ratio above 10$^{-4}$ at CMB scales, which is two orders
of magnitude below the published goal of a tensor-to-scalar ratio of $10^{-2}$ by
the planned next generation CMB polarization space mission by NASA ({\it CMBpol}).  

A less sensitive version of BBO (BBO-lite) with a sensitivity to gravitational waves $\Omega_{\rm GW}h^2$ above $10^{-15}$
is unlikely to be useful given that the predictions suggest backgrounds well below this level when the
tensor-to-scalar ratio is below 0.3. Even if the tensor-to-scalar ratio were to be above 10$^{-2}$, 
we suggest that BBO-grand will be useful to study inflationary models 
as standard BBO will only allow a marginal detection of the amplitude while leaving the spectral index unconstrained.  
The polarized foregrounds,  
such as dust and synchrotron, are expected to limit CMB studies to a tensor-to-scalar a few times 10$^{-4}$, which is
well above the ultimate limit of cosmic shear. Unless direct detection experiments are
also affected by foreground gravitational wave sources, laser interferometers may allow us to expand the search for gravitational
waves down to a tensor-to-scalar ratio of 10$^{-6}$. 

Based on our models, we find that simple power-law approximations are not adequate to extrapolate either information at CMB scales
to direct detection experiments or vice-versa. The differences are largely understood through scale dependence of the
tensor tilt produced in the slow-roll inflationary models through flow equations. We also show that the simple
assumption of $n_T(k_{\rm CMB})\approx n_T({\rm 0.1 Hz})$ cannot be used to establish inflationary consistency relations
by combining measurements at CMB and direct detection scales. We advocate additional measurements of the primordial scalar power spectrum
at large wavenumbers beyond CMB as a way to improve model selection and predictions of inflation.

\section{Acknowledgments}
AC and BF are supported by DOE and AM is supported by MURST through COFIN contract no. 2004027755. 
AC thanks members of the {\it Dipartimento di Fisica} and {\it INFN} at 
Universita' di Roma for hospitality while this work was
initiated and participants of the ``Inflation+25'' conference at {\it Institute d'Astrophysique}, Paris
for comments and suggestions that initiated this work.  We thank Will Kinney, Tristan Smith, Naoki Seto, 
and Manoj Kaplinghat for useful discussions related to this work.


\begin{thebibliography}{99}
\frenchspacing


\bibitem{Spergel}
  D.~N.~Spergel {\it et al.}  [WMAP Collaboration],
  %``First Year Wilkinson Microwave Anisotropy Probe (WMAP) Observations:
  %Determination of Cosmological Parameters,''
  Astrophys.\ J.\ Suppl.\  {\bf 148}, 175 (2003);
  %%CITATION = ASTRO-PH 0302209;%%
 D.~N.~Spergel {\it et al.},
  %``Wilkinson Microwave Anisotropy Probe (WMAP) three year results:
  %Implications for cosmology,''
  arXiv:astro-ph/0603449.
  %%CITATION = ASTRO-PH 0603449;%%
  H.~V.~Peiris {\it et al.},
  %``First year Wilkinson Microwave Anisotropy Probe (WMAP) observations:
  %Implications for inflation,''
  Astrophys.\ J.\ Suppl.\  {\bf 148}, 213 (2003).
  %%CITATION = ASTRO-PH 0302225;%%

\bibitem{Guth81}
  A.~H.~Guth, \PRD\ {\bf 23}, 347 (1981); A.~Albrect and P.~J.~Steinhardt, \PRL\ {\bf 48} 1220 (1982);
  A.~D.~Linde, Phys.\ Lett. B{\bf 108}, 389 (1982).



\bibitem{Abbott84}
  A.~Starobinskii, JETP\ Lett {\bf 30}, 682 (1979);
  L.~F.~Abbott and M.~B.~Wise, Nucl.\ Phys.\ {\bf B244}, 541 (1984); A.~Starobinskii, Sov.\ Astron.\ Lett {\bf 11}, 133 (1985);
V.~A.~Rubakov, M.~V.~Sazhin and A.~V.~Veryaskin,
  %``Graviton Creation In The Inflationary Universe And The Grand Unification
  %Scale,''
  Phys.\ Lett.\ B {\bf 115}, 189 (1982);
  %%CITATION = PHLTA,B115,189;%%
  R.~Fabbri and M.~d.~Pollock,
  % ``The Effect Of Primordially Produced Gravitons Upon The Anisotropy Of The
  %Cosmological Microwave Background Radiation,''
  Phys.\ Lett.\ B {\bf 125}, 445 (1983);
  %%CITATION = PHLTA,B125,445;%%
V.~Sahni, \PRD\ {\bf 42}, 453 (1990);
B.~Allen, \PRD\ {\bf 37}, 2078 (1988).


\bibitem{Kamionkowski:96}
  M.~Kamionkowski, A.~Kosowsky and A.~Stebbins,
  %``A probe of primordial gravity waves and vorticity,''
  Phys.\ Rev.\ Lett.\  {\bf 78}, 2058 (1997);
  %%CITATION = ASTRO-PH 9609132;%%
  U.~Seljak and M.~Zaldarriaga,
  %``Signature of gravity waves in polarization of the microwave background,''
  Phys.\ Rev.\ Lett.\  {\bf 78}, 2054 (1997).
  %%CITATION = ASTRO-PH 9609169;%%


\bibitem{Hinshaw:2006ia}
  G.~Hinshaw {\it et al.},
   ``Three-year Wilkinson Microwave Anisotropy Probe (WMAP) observations:
  %Temperature analysis,''
  arXiv:astro-ph/0603451.
  %%CITATION = ASTRO-PH 0603451;%%


\bibitem{Kinney:06}
 W.~H.~Kinney, E.~W.~Kolb, A.~Melchiorri and A.~Riotto,
   ``Inflation model constraints from the Wilkinson microwave anisotropy probe
  %three-year data,''
  Phys.\ Rev.\ D {\bf 74}, 023502 (2006)
  [arXiv:astro-ph/0605338].
  %%CITATION = ASTRO-PH 0605338;%%


\bibitem{Weiss}
R.~Weiss et al. CMB Task Force Report, www.science.doe.gov/hep/TFCRreport.pdf

\bibitem{foregrounds}
 M.~Amarie, C.~Hirata and U.~Seljak,
  %``Detectability of tensor modes in the presence of foregrounds,''
  Phys.\ Rev.\ D {\bf 72}, 123006 (2005)
  [arXiv:astro-ph/0508293];
  %%CITATION = ASTRO-PH 0508293;%%
   E.~Carretti, G.~Bernardi and S.~Cortiglioni,
  %``B-Mode contamination by synchrotron emission from 3-years WMAP data,''
  arXiv:astro-ph/0609288;
  %%CITATION = ASTRO-PH 0609288;%%
  A.~Amblard, A.~Cooray and M.~Kaplinghat, in preparation.

\bibitem{Kesden}
L.~Knox and Y.S.~Song,  Phys.\ Rev.\ Lett. {\bf 89}, 011303 (2002);
 M.~Kesden, A.~Cooray and M.~Kamionkowski,
  %``Separation of gravitational-wave and cosmic-shear contributions to  cosmic
  %microwave background polarization,''
  Phys.\ Rev.\ Lett.\  {\bf 89}, 011304 (2002)
  [arXiv:astro-ph/0202434];
  %%CITATION = ASTRO-PH 0202434;%%
U.~Seljak and C.~M.~Hirata,
  %``Gravitational lensing as a contaminant of the gravity wave signal in
  %CMB,''
  Phys.\ Rev.\ D {\bf 69}, 043005 (2004)
  [arXiv:astro-ph/0310163].
  %%CITATION = ASTRO-PH 0310163;%%

\bibitem{BBO}
E.~S.~Phinney et al. The Big Bang Observer, NASA Mission Concept Study (2003); URL http://universe.nasa.gov/program/bbo.html

\bibitem{Seto:01}
  N.~Seto, S.~Kawamura and T.~Nakamura,
  %``Possibility of direct measurement of the acceleration of the universe
  %using 0.1-Hz band laser interferometer gravitational wave antenna in
  %space,''
  Phys.\ Rev.\ Lett.\  {\bf 87}, 221103 (2001).
  %%CITATION = ASTRO-PH 0108011;%%

\bibitem{Crowder}
  J.~Crowder and N.~J.~Cornish,
  %``Beyond LISA: Exploring future gravitational wave missions,''
  Phys.\ Rev.\ D {\bf 72}, 083005 (2005)
  [arXiv:gr-qc/0506015];
  %%CITATION = GR-QC 0506015;%%
C.~Ungarelli and A.~Vecchio,
  %``High energy physics and the very-early universe with LISA,''
  Phys.\ Rev.\ D {\bf 63}, 064030 (2001)
  [arXiv:gr-qc/0003021].
  %%CITATION = GR-QC 0003021;%%

\bibitem{Turner:96}
  M.~S.~Turner,
  %``Detectability of inflation-produced gravitational waves,''
  Phys.\ Rev.\ D {\bf 55}, 435 (1997);
  %%CITATION = ASTRO-PH 9607066;%%

\bibitem{Ungarelli:05}
C.~Ungarelli, P.~Corasaniti, R.~A.~Mercer and A.~Vecchio,
  %``Gravitational waves, inflation and the cosmic microwave background: Towards
  %testing the slow-roll paradigm,''
  Class.\ Quant.\ Grav.\  {\bf 22}, S955 (2005).
  %%CITATION = ASTRO-PH 0504294;%%

\bibitem{Smith:06}
  T.~L.~Smith, M.~Kamionkowski and A.~Cooray,
  %``Direct detection of the inflationary gravitational wave background,''
  arXiv:astro-ph/0506422.
  %%CITATION = ASTRO-PH 0506422;%%

\bibitem{Smith}
 T.~L.~Smith, H.~V.~Peiris and A.~Cooray,
  %``Deciphering inflation with gravitational waves: Cosmic microwave
  %background polarization vs. direct detection with laser  interferometers,''
  Phys.\ Rev.\ D {\bf 73}, 123503 (2006)
  [arXiv:astro-ph/0602137].
  %%CITATION = ASTRO-PH 0602137;%%

\bibitem{Efstathiou}
 S.~Chongchitnan and G.~Efstathiou,
  %``Prospects for direct detection of primordial gravitational waves,''
  Phys.\ Rev.\ D {\bf 73}, 083511 (2006)
  [arXiv:astro-ph/0602594];
  %%CITATION = ASTRO-PH 0602594;%%
G.~Efstathiou and S.~Chongchitnan,
  %``The search for primordial tensor modes,''
  arXiv:astro-ph/0603118.
  %%CITATION = ASTRO-PH 0603118;%%

\bibitem{Coo:05}
  A.~Cooray,
  %``Primordial gravitational waves and inflation: CMB and direct detection
  %with space-based laser interferometers,''
  Mod.\ Phys.\ Lett.\ A {\bf 20},  (19??)
  [arXiv:astro-ph/0503118].
  %%CITATION = ASTRO-PH 0503118;%%



\bibitem{Turner:93}
  M.~S.~Turner, M.~J.~White and J.~E.~Lidsey,
  %``Tensor perturbations in inflationary models as a probe of cosmology,''
  Phys.\ Rev.\ D {\bf 48}, 4613 (1993)
  [arXiv:astro-ph/9306029];
  %%CITATION = ASTRO-PH 9306029;%%
J.~R.~Pritchard and M.~Kamionkowski,
  %``Cosmic microwave background fluctuations from gravitational waves: An
  %analytic approach,''
  Annals Phys.\  {\bf 318}, 2 (2005)
  [arXiv:astro-ph/0412581];
  %%CITATION = ASTRO-PH 0412581;%%
Y.~Watanabe and E.~Komatsu,
%   ``Improved calculation of the primordial gravitational wave spectrum in the
  %standard model,''
  Phys.\ Rev.\ D {\bf 73}, 123515 (2006)
  [arXiv:astro-ph/0604176].
  %%CITATION = ASTRO-PH 0604176;%%


%\cite{hoffman/turner:2001}
\bibitem{hoffman/turner:2001}
  M.~B.~Hoffman and M.~S.~Turner,
  %``Kinematic constraints to the key inflationary observables,''
  Phys.\ Rev.\ D {\bf 64}, 023506 (2001)
  [arXiv:astro-ph/0006321].
  %%CITATION = ASTRO-PH 0006321;%%

%\cite{kinney:2002}
\bibitem{kinney:2002}
  W.~H.~Kinney,
  %``Inflation: Flow, fixed points and observables to arbitrary order in  slow
  %roll,''
  Phys.\ Rev.\ D {\bf 66}, 083508 (2002).
  %%CITATION = ASTRO-PH 0206032;%%

 %\cite{easther/kinney:2003}
\bibitem{easther/kinney:2003}
  R.~Easther and W.~H.~Kinney,
  %``Monte Carlo reconstruction of the inflationary potential,''
  Phys.\ Rev.\ D {\bf 67}, 043511 (2003)
  [arXiv:astro-ph/0210345].
  %%CITATION = ASTRO-PH 0210345;%%

%%\cite{lidsey/etal:1995}
\bibitem{Lidsey}
  J.~E.~Lidsey, A.~R.~Liddle, E.~W.~Kolb, E.~J.~Copeland, T.~Barreiro and M.~Abney,
  %``Reconstructing the inflaton potential: An overview,''
  Rev.\ Mod.\ Phys.\  {\bf 69}, 373 (1997)
  [arXiv:astro-ph/9508078];
  %%CITATION = ASTRO-PH 9508078;%%
  L.~P.~Grishchuk and Yu.~V.~Sidorav, in {\it Fourth Seminar on Quantum Gravity}. eds. M.~A.~Markov, V.~A.~Berezin and V.~P.~Frolov (World Scientific: Singapore, 1988);
  A.~G.~Muslimov, Class. Quant. Hrav. {\bf 7}, 231 (1990);
  D.~S.~Sapolek and J.~R.~Bond, \PRD\ {\bf 42}, 3936 (1990).

%\cite{Liddle:1994dx}
\bibitem{Liddle}
  A.~R.~Liddle, P.~Parsons and J.~D.~Barrow,
  %``Formalizing the slow roll approximation in inflation,''
  Phys.\ Rev.\ D {\bf 50}, 7222 (1994)
  [arXiv:astro-ph/9408015].
  %%CITATION = ASTRO-PH 9408015;%%

%\cite{liddle:2003}
\bibitem{liddle:2003}
  A.~R.~Liddle,
  %``On the inflationary flow equations,''
  Phys.\ Rev.\ D {\bf 68}, 103504 (2003).
  %%CITATION = ASTRO-PH 0307286;%%

\bibitem{Seto:06}
  N.~Seto,
  %``Correlation analysis of stochastic gravitational wave background around
  %0.1-Hz - 1-Hz,''
  arXiv:gr-qc/0510067;
  %%CITATION = GR-QC 0510067;%%
  H.~Kudoh, A.~Taruya, T.~Hiramatsu, and Y.~Himemoto, gr-qc/0511145.

\bibitem{Song}
Y.-S.~Song and L.~Knox (2003), astro-ph/0305411.


\bibitem{Boyle}
L.~A.~Boyle and P.~J.~Steinhardt,
  %``Probing the early universe with inflationary gravitational waves,''
  arXiv:astro-ph/0512014.
  %%CITATION = ASTRO-PH 0512014;%%


\bibitem{Weinberg:04}
 S.~Weinberg, \PRD\ {\bf 69}, 023503 (2004); S.~Bashinsky, astro-ph/0505502 (2005).



%\cite{peiris/etal:2003}
%\bibitem{peiris/etal:2003}
%  H.~V.~Peiris {\it et al.},
%  %``First year Wilkinson Microwave Anisotropy Probe (WMAP) observations:
%  %Implications for inflation,''
%  Astrophys.\ J.\ Suppl.\  {\bf 148}, 213 (2003)
%  [arXiv:astro-ph/0302225].
%  %%CITATION = ASTRO-PH 0302225;%%

\bibitem{Alabidi}
  L.~Alabidi and D.~H.~Lyth,
  %``Inflation models after WMAP year three,''
  JCAP {\bf 0608}, 013 (2006)
  [arXiv:astro-ph/0603539].
  %%CITATION = ASTRO-PH 0603539;%%


%\cite{Dodelson:2003vq}
\bibitem{Dodelson}
  S.~Dodelson and L.~Hui,
  %``A horizon ratio bound for inflationary fluctuations,''
  Phys.\ Rev.\ Lett.\  {\bf 91}, 131301 (2003)
  [arXiv:astro-ph/0305113].
  %%CITATION = ASTRO-PH 0305113;%%



\bibitem{PeiEas06}
  H.~Peiris and R.~Easther,
   ``Slow roll reconstruction: Constraints on inflation from the 3 year WMAP
  %dataset,''
  arXiv:astro-ph/0609003.
  %%CITATION = ASTRO-PH 0609003;%%

\bibitem{EfsMac05}
  G.~Efstathiou and K.~J.~Mack,
  %``The Lyth Bound Revisited,''
  JCAP {\bf 0505}, 008 (2005)
  [arXiv:astro-ph/0503360].
  %%CITATION = ASTRO-PH 0503360;%%








 
%\cite{Stewart:1993bc}
\bibitem{stewart/lyth:1993}
  E.~D.~Stewart and D.~H.~Lyth,
  %``A More accurate analytic calculation of the spectrum of cosmological
  %perturbations produced during inflation,''
  Phys.\ Lett.\ B {\bf 302}, 171 (1993)
  [arXiv:gr-qc/9302019].
  %%CITATION = GR-QC 9302019;%%


\bibitem{HuOka02}
     W.~Hu and T.~Okamoto, \apj\ {\bf 574}, 566 (2002)
     [arXiv:astro-ph/0111606];
M.~Kesden, A.~Cooray, and M.~Kamionkowski,
        \PRD\ {\bf 67}, 123507 (2003) [arXiv:astro-ph/0302536];
C.~M.~Hirata and U.~Seljak,
%``Reconstruction of lensing from the cosmic microwave background
%polarization,''
Phys.\ Rev.\ D {\bf 68}, 083002 (2003)
[arXiv:astro-ph/0306354].
%%CITATION = ASTRO-PH 0306354;%%


\end{thebibliography}
\end{document}